\documentclass[12pt]{article}

\pdfoutput=1

\usepackage[top=80pt,bottom=85pt,left=58pt,right=58pt]{geometry}
\usepackage{amssymb,amsmath,xypic}
\usepackage{slashed}
\usepackage{graphicx,color,subfigure}
\usepackage{float}
\usepackage{sectsty}
\usepackage{setspace}
\usepackage{cite}
\usepackage{empheq}
\usepackage{graphicx,wrapfig,lipsum}

\usepackage[debug,pageanchor=false]{hyperref}
\definecolor{link}{rgb}{.8,.15,.1}
\hypersetup{colorlinks=true,linkcolor=link,citecolor=link,urlcolor=link,linktocpage}

 \allowdisplaybreaks

\makeatletter
\@addtoreset{equation}{section}
\makeatother

\newcommand*\widefbox[1]{\fbox{\hspace{2em}#1\hspace{2em}}}

\newcommand{\beq}{\begin{equation}}
\newcommand{\eeq}{\end{equation}}
\newcommand{\bea}{\begin{eqnarray}}
\newcommand{\eea}{\end{eqnarray}}
\newcommand{\nn}{\nonumber}

\newcommand{\e}{\text{e}}

\begin{document}

\begin{titlepage}

\begin{center}

\vskip .5in %.3in 
\noindent

{\Large \bf{G-structures for black hole near-horizon geometries}}

		\bigskip\medskip
  Andrea Legramandi$^{a}$\footnote{andrea.legramandi@unitn.it}, Niall T. Macpherson$^{b}$\footnote{macphersonniall@uniovi.es}, Achilleas Passias$^{c}$\footnote{achilleas.passias@lpthe.jussieu.fr} \\

\bigskip\medskip
{\small 

	$a$: Pitaevskii BEC Center, CNR-INO and Dipartimento di Fisica, Universit\'a di Trento, I38123 Trento, Italy  \\
	and
	\\
	INFN-TIFPA, Trento Institute for Fundamental Physics and Applications, Trento, Italy \vskip 3mm
	
	$b$: Department of Physics, University of Oviedo,\\ 
	Avda. Federico Garcia Lorca s/n, 33007 Oviedo
\\
	and
\\
	 Instituto Universitario de Ciencias y Tecnolog\'ias Espaciales de Asturias (ICTEA),\\ 
	Calle de la Independencia 13, 33004 Oviedo, Spain }\vskip 3mm
	
  $c$:	Sorbonne Universit\'e, UPMC Paris 06, UMR 7589, LPTHE,
75005 Paris, France

		\vskip 1.8cm %.6cm
		\textbf{ Abstract}\\
			\vskip .2cm %.6cm
\end{center}
We derive necessary and sufficient conditions for warped AdS$_2$ solutions of Type II supergravity to preserve ${\cal N}=1$ supersymmetry, in terms of bispinors. Such solutions generically support an SU$(3)$-structure on their internal manifold M$_8$,  which can experience an enhancement to a G$_2$-structure. We perform an SU$(3)$-structure torsion classes analysis and express the fluxes and other physical fields in terms of these, in general. We use our results to derive two new classes of AdS$_2$ solutions. In (massive) Type IIA supergravity we derive an ${\mathcal N}=1$ supersymmetric class for which M$_8$ decomposes as a weak G$_2$-manifold foliated over an interval and which is locally defined in terms of a degree three polynomial. In Type IIB supergravity we find a class of AdS$_2\times\text{S}^2\times\text{CY}_2\times\Sigma_2$ solutions preserving small ${\cal N}=4$ supersymmetry, governed by a harmonic function on $\Sigma_2$ and partial differential equations reminiscent of D3-D7-brane configurations.

		\vskip .1in
	
	\noindent 

\vskip .1in

\noindent

\noindent

\vfill
\eject

\end{titlepage}

\tableofcontents

\setlength{\parskip}{1em}
\section{Introduction}
Black hole mechanics has been a driving force and guide in the quest of a quantum theory of gravity. Quantum mechanically black holes behave as thermodynamic objects with temperature and entropy. The Bekenstein--Hawking formula expresses the entropy of a black hole in terms of the area of its event horizon, and the microscopic origin of the entropy is a problem that every quantum theory of gravity aspires to address. Extremal black holes play a prominent role as their quantum description is under better control.

All known supersymmetric extremal black holes possess an AdS$_2$ factor in their near-horizon geometry and one expects that the holographic correspondence between two-dimensional anti-de Sitter space (AdS$_2$) and superconformal quantum mechanics (SCQM) is of value in their study. The AdS$_2$/SCQM correspondence is less understood than its higher-dimensional counterparts owning to the special features of gravity in spacetimes that asymptote to AdS$_2$. Identifying AdS$_2$/SCQM pairs in string theory is thus desirable and motivates the analysis of the space of AdS$_2$ solutions. 

Some interesting existing examples of AdS$_2$ solutions and their physical relevance are the following.  There is the GK solution \cite{Gauntlett:2007ts,Donos:2008ug} which appears prominently in the  holographic dual to $\mathcal{I}$-extremisation \cite{Couzens:2018wnk, Gauntlett:2019roi}. Various compactifications of higher-dimensional AdS solutions,  dual to likewise compactified conformal field theories (CFTs), have been constructed; see for instance \cite{Couzens:2021rlk,Ferrero:2021ovq,Suh:2021hef,
Hosseini:2017fjo} for compactifications on  Riemann surfaces and \cite{Suh:2018szn,Hosseini:2018usu,Legramandi:2021aqv,Couzens:2022lvg,Faedo:2022rqx} for compactifications on higher-dimensional spaces. There are also various holographic duals to defects in higher-dimensional CFTs \cite{DHoker:2007mci,Chiodaroli:2009xh,Chiodaroli:2009yw,Dibitetto:2018gtk,Lozano:2020sae,Lozano:2021fkk,Lozano:2022vsv,Lozano:2022swp} exhibiting AdS$_2$ factors.

AdS$_2$ solutions in Type II supergravity can often be generated from existing AdS$_3$ solutions via Hopf fiber T-duality or Hopf fiber reduction from $d=11$ supergravity \cite{Boonstra:1998yu}. This has been exploited in recent works to yield the first examples of new classes of supersymmetic AdS$_2$ solutions whose dual quantum mechanics where explored in \cite{Lozano:2020sae,Lozano:2020txg,Lozano:2021rmk,Lozano:2021fkk,Lozano:2022vsv,Lozano:2022swp,Rigatos:2022ktp}, see also \cite{Dibitetto:2019nyz}. Another way to generate AdS$_2$ solutions is via SL$(2)$ non-Abelian T-duality \cite{Ramirez:2021tkd,Lozano:2021rmk,Conti:2023naw,Conti:2023rul}.

Several classifications of maximally supersymmetric AdS$_2$ solutions already exist \cite{DHoker:2007mci,Corbino:2017tfl,Corbino:2018fwb,Dibitetto:2018gbk,Corbino:2020lzq},  though the ${\cal N}=(8,0)$ AdS$_3$ classification of \cite{Legramandi:2020txf}  suggests that likely, this only scratches the surface of such possibilities. Additionally, in \cite{Hong:2019wyi}, minimally supersymmetric AdS$_2$ solutions of $d=11$ supergravity were classified under the assumption that they preserve an SU$(4)$-structure. We prove here that AdS$_2$ solutions not conforming to this assumption merely embed AdS$_2$ inside higher-dimensional AdS spaces, so \cite{Hong:2019wyi} is actually general. See also \cite{Kim:2006qu,MacConamhna:2006nb} for  earlier partial classification. 

In this paper we aim for a classification of supersymmetric AdS$_2$ solutions of Type II supergravity, leveraging the techniques involving bispinors and G-structures, that have been proven effective in the classification of other supersymmetric solutions.  Supersymmetric AdS$_2$ solutions can support a wide variety of superconformal algebras. Those that can be embedded into $d=10$ and $d=11$ supergravities were classified in \cite{Gran:2017qus}. The possible algebras are actually the same as the (simple) chiral superconformal  algebras that CFTs in $d=2$ can support; see for instance \cite{Fradkin:1992bz}. Like their AdS$_3$ counterparts, sixteen real supercharges, so ${\cal N}=8$, is maximal for solutions containing an AdS$_2$ factor. Our focus here will be on solutions that preserve at least minimal supersymmetry, i.e.\ ${\cal N}=1$, which is two real supercharges.  

The layout of the paper is as follows:

In section \ref{sec:SUSYads2} we present necessary and sufficient conditions for AdS$_2$ solutions of Type II supergravity to preserve ${\cal N}=1$ supersymmetry, under the assumption that they do not experience an enhancement to a higher-dimensional AdS space, as they do generically. We then move on to classify these conditions in terms of G-structures. We find that the internal 8-manifold M$_8$ generically supports an SU$(3)$-structure, though a limit exists where this gets enhanced to a G$_2$-structure. This section is supplemented by the technical appendix \ref{sec:derivation}, where many of these results are derived.  

In section \ref{sec:recoveringAdS} we perform a non-trivial check of the conditions for supersymmetry derived earlier. All supersymmetric AdS$_3$ solutions of $d=11$ supergravity and  of Type II supergravity with purely magnetic NSNS flux, can be mapped to supersymmetric AdS$_2$ solutions by either dimensionally reducing or T-dualising on the Hopf fiber of AdS$_3$. Such AdS$_3$ solutions have been classified in the literature in terms of G-structures: see \cite{Martelli:2003ki,Tsimpis:2005kj,Legramandi:2020txf} for $d=11$  and  \cite{Dibitetto:2018ftj,Passias:2019rga,Passias:2020ubv,Macpherson:2021lbr} for $d=10$. Our AdS$_2$ conditions should then reproduce those of these AdS$_3$ classifications in certain limits and we show this is indeed the case.

Section \ref{sec:torsionclasses} contains a general classification of the possible types of solutions in terms of SU$(3)$-structure torsion classes. It turns out that the possible internal manifolds and fluxes these solutions can support is quite broad. This is perhaps unsurprising as the internal space is large and the supersymmetry low. Nonetheless, this section will serve as a good road map of which backgrounds can be obtained  for supersymmetric AdS$_2$.

The main text ends with section \ref{eq:examples}, where, as a display of the utility of the tools we provide in this work, we derive two new and non-trivial classes of AdS$_2$ solutions, one in each of Type IIA and IIB supergravities. The first, in section \ref{sec:IIAexample}, is a generalisation of a class of ${\cal N}=8$ solutions in massive Type IIA supergravity, with AdS$_2 \times $S$^7$ foliated over an interval \cite{Dibitetto:2018gbk}. We generalise by replacing S$^7$ with an arbitrary weak G$_2$-manifold, allowing the fluxes to depend on its associated G$_2$-structure. The result is a class of ${\cal N}=1$ solutions governed by two ordinary differential equations, which in two distinct cases can be solved in terms of a degree three polynomial. In section \ref{sec:IIBexample}, we derive a class of solutions on a warped product of $\text{AdS}_2\times\text{S}^2\times \text{CY}_2\times \Sigma_2$, governed by a harmonic function on $\Sigma_2$ and a system of partial differential equations that generalises those of localised D3-branes with D7-branes, that have CY$_2$ as their relative codimensions. It admits a limit in which the near-horizon of a $d=4$ black hole is recovered, so it might be useful for the study of microstate counting in the future. This class generalises a class of solutions generated via T-duality in \cite{Lozano:2021rmk}, and also has some partial intersection with the classification of \cite{Chiodaroli:2009yw,Chiodaroli:2009xh}.

Finally our work is supplemented by several in depth appendices. Our conventions can be found in appendix \ref{sec:convensions}. In appendices \ref{sec:appenixAdS2bilinears} and \ref{sec:refinment} we lay the groundwork for the following appendix, by analysing Killing spinor bilinears on AdS$_2$ and refining (for generic spacetimes) the pairing constraints of \cite{Tomasiello:2011eb}, for scenarios where the $d=10$  Killing spinors define a time-like Killing vector. As previously mentioned, appendix \ref{sec:derivation} derives most of the results of section \ref{sec:SUSYads2}. Finally, in appendix \ref{sec:comment}, we prove that the classification of AdS$_2$ solutions in \cite{Hong:2019wyi} is actually fully general if one insists that they are not merely the embedding of AdS$_2$ into some higher-dimensional AdS space.

\section{\texorpdfstring{Geometric conditions for ${\cal N}=1$ supersymmetric AdS$_2$ in Type II supergravity}{Geometric conditions for ${\cal N}=1$ supersymmetric AdS(2) in Type II supergravity}}\label{sec:SUSYads2}
In this section we present a set of necessary and sufficient geometric conditions for an AdS$_2$ solution of Type II supergravity to preserve minimal supersymmetry. This section is largely a summary of appendix \ref{sec:derivation}, where these conditions are derived.

An AdS$_2$ solution of Type II supergravity has bosonic field content that can, by definition, be decomposed as
\begin{align}
ds^2&=e^{2A}ds^2(\text{AdS}_2)+ds^2(\text{M}_8),\nn\\
F&= f_{\pm}+e^{2A}\text{vol}(\text{AdS}_2)\wedge\star_8 \lambda(f_{\pm}),~~~~ H=e^{2A}\text{vol}(\text{AdS}_2)\wedge H_1+H_3,\label{eq:AdS2vacua} 
\end{align}
where $H$ is the NSNS 3-form flux and $F$ is the $d=10$ RR polyform flux, with the upper/lower signs taken in Type IIA/IIB throughout, with $\pm$ labelling even/odd form degree when appearing on forms or chirality when appearing on spinors. The AdS$_2$ warp factor $e^{2A}$, dilaton $\Phi$ and the forms $(f_{\pm},H_1,H_3)$ have support on the internal manifold M$_8$ alone, and the operator $\lambda$ acts on a $k$-form as $\lambda(C_k)=(-1)^{\lfloor\frac{k}{2}\rfloor}C_k$. The fluxes, away from the loci of sources, should obey the following magnetic
\beq\label{eq:magBIS}
dH_3=0,~~~~ d_{H_3} f_{\pm}= 0
\eeq
and the following electric
\beq\label{eq:elBIS}
d(e^{2A}H_1)=0,~~~~ d_{H_3} (e^{2A}\star_8\lambda(f_{\pm}))= e^{2A}H_1\wedge f_{\pm},
\eeq 
Bianchi identities. Further details of our conventions can be found in appendix \ref{sec:convensions}.

When an AdS$_2$ solution preserves supersymmetry it does so in terms of the Killing spinors of AdS$_2$. These come in two chiral variants $(\zeta_+,\zeta_-)$, which can be taken to be Majorana without loss of generality, and obey the Killing spinor equations
\beq
\nabla_{\mu}^{\text{AdS}_2}\zeta_+= \frac{m}{2}\gamma^{(2)}_{\mu}\zeta_-,~~~~ \nabla_{\mu}^{\text{AdS}_2}\zeta_-= \frac{m}{2}\gamma^{(2)}_{\mu}\zeta_+,
\eeq  
where $m$ is the inverse AdS$_2$ radius; more details are given in appendix \ref{sec:appenixAdS2bilinears}.  In $d=10$ each of these couple to Majorana--Weyl spinors on M$_8$  $(\chi^i_+,\chi^i_-)$ for $i=1,2$ such that the $d=10$ Majorana--Weyl spinors decompose as\footnote{We are assuming a factorization of the $d=10$ spinors.}
\beq
\epsilon_{1}=\zeta_+\otimes\chi_{1+}+\zeta_-\otimes\chi_{1-},~~~~ \epsilon_{2}=\zeta_+\otimes\chi_{2\mp}+\zeta_-\otimes\chi_{2\pm},
\eeq
where none of the $d=8$ spinors can be zero.\footnote{See the discussion around  \eqref{eq:nozerointernalspinors}.} This decomposition preserves 2 out of the 32 supersymmetries in ten dimensions. Going forward, we find it helpful to define the non-chiral $d=8$ spinors
\beq
\chi_1:= \chi_{1+}+\chi_{1-},~~~~ \chi_2:= \chi_{2+}+ \chi_{2-}.
\eeq
From this staring point, in appendix \ref{sec:derivation}, we are able to derive necessary and sufficient geometric conditions for a solution of the form \eqref{eq:AdS2vacua} to preserve minimal supersymmetry. We will now summarise these conditions.

The first thing to appreciate is that totally generic AdS$_2$ solutions  experience an enhancement to AdS$_3$, unless one imposes that
\beq\label{eq:noads3cond}
\chi^{\dag}_{1}\gamma_a \chi_{1}=\pm \chi^{\dag}_{2}\gamma_a \chi_{2},~~~~\chi_1^{\dag}\hat\gamma \chi_1=\pm\chi_2^{\dag}\hat\gamma \chi_2,~~~~ |\chi_1|^2=|\chi_2|^2= c e^{A},
\eeq
where $\gamma_a$ are a basis of gamma matrices on M$_8$, $\hat\gamma$ the corresponding chirality matrix and $c$ is an integration constant. Thus, if one is interested in true AdS$_2$ solutions, \eqref{eq:noads3cond} needs to be imposed.\footnote{Notice that imposing this makes our Ansatz inconsistent with all AdS$_d$ for $d>2$, not only AdS$_3$, as they can all be expressed in terms of an AdS$_3$ factor.} Given this, solutions can be defined in terms of the following 0- and 1-form spinor bilinears
\beq
c e^{A}\cos\beta :=\chi_1^{\dag}\hat\gamma \chi_1=\pm\chi_2^{\dag}\hat\gamma \chi_2,~~~~c e^{A}\sin\beta V :=\chi^{\dag}_{1}\gamma_a \chi_{1} \e^a=\pm \chi^{\dag}_{2}\gamma_a \chi_{2} \e^a,~~~~ V\cdot V=1, \label{eq:noads3conds}
\eeq
for $\e^a$ a vielbein  basis on M$_8$ and where $\sin\beta = 0$ is incompatible with $m\neq 0$ (for $m=0$ AdS$_2$ blows up to $\text{Mink}_2$). In addition to this, we need to define the polyform bilinears\footnote{Strictly speaking the left-hand side of these expressions is not a polyform, rather it is the components of a polyform whose indices have be contracted with an appropriate number of antisymmetric products of gamma matrices. However such objects can be mapped to forms under the Clifford map, i.e.\ $\slashed{\psi}=\chi_1\otimes \chi_2^{\dag}=\frac{1}{16}\sum_{n=0}^8\frac{1}{n!}\chi_2^{\dag}\gamma_{\underline{a}_n\dots\underline{a}_1}\chi_1 \gamma^{\underline{a}_1\dots\underline{a}_n}$ and $\psi=\frac{1}{16}\sum_{n=0}^8\frac{1}{n!}\chi_2^{\dag}\gamma_{\underline{a}_n\dots\underline{a}_1}\chi_1 \e^{\underline{a}_1\dots\underline{a}_n}$ are equivalent objects. We are simply suppressing the Dirac slash in the above.}
\beq 
\psi:= \chi_1\otimes \chi_2^{\dag},~~~~\hat{\psi}:= \hat\gamma\chi_1\otimes \chi_2^{\dag},
\eeq
with odd/even form degree components $(\psi_{\mp},\hat\psi_{\mp})$. In terms of these, necessary and sufficient conditions for supersymmetry (when $m\neq 0$) are given by
\begin{subequations}
\begin{empheq}[box=\widefbox]{align}
&e^{2A}H_1=me^{A} \sin\beta V-d(e^{2A}\cos\beta),~~~~d(e^{A}\sin\beta V)=0,\label{BPS1}\\
&d_{H_3}(e^{-\Phi}\psi_{\pm})=\pm \frac{c}{16}e^{A}\sin\beta V\wedge f_{\pm},\label{BPS2}\\
&d_{H_3}(e^{A-\Phi}\hat\psi_{\mp})- m e^{-\Phi}\psi_{\pm}=\mp\frac{c }{16}e^{2A}(\star_8\lambda f_{\pm}+\cos\beta f_{\pm}),\label{BPS3}\\
&(\psi_{\pm},f_{\pm})_8=\pm\frac{c}{4} e^{-\Phi}\left(m-\frac{1}{2}e^A\sin\beta \iota_{V} H_1\right)\text{vol}(\text{M}_8).\label{BPS8dpairing}
\end{empheq}
\end{subequations}
Here the final condition is a pairing constraint where in general the bracket defines the $k$-dimensional Chevalley--Mukai pairing  $(X, Y)_k:=  X\wedge \lambda(Y)\lvert_k$ (where $_k$ denotes the $k$-form contribution). In the appendix, three further pairing constraints are presented 
\begin{subequations}
\begin{align}
&(\psi_{\mp}, f_{\pm})_7=0,\label{BPS4s}\\
&(\hat\psi_{\mp}, f_{\pm})_7=\pm \frac{1}{8}e^{A-\Phi}c\star_8( 2dA+\cos\beta H_1),\label{BPS5}\\
&(\hat\psi_{\mp}, \star\lambda f_{\pm })_7=\pm \frac{1}{8}e^{A-\Phi}c\star_8( 2 \cos\beta dA+ H_1-2 e^{-A} m \sin\beta V)\label{BPS6},
\end{align}
\end{subequations}
where it is found that these actually imply \eqref{BPS8dpairing}. However, we establish that \eqref{BPS4s}-\eqref{BPS6}  are implied by \eqref{BPS1}-\eqref{BPS8dpairing}  during our torsion classes analysis in section \ref{sec:torsionclasses}. 
The above conditions imply several others, for instance one can derive the following condition independent of  $f_{\pm}$
\beq\label{BPS:decendent}
d_{H_3}(e^{A-\Phi}\psi_{\mp})=0,
\eeq
which follows from $\eqref{BPS2}\wedge V$ and can be useful for extracting necessary conditions.  

Of course supersymmetry alone is not sufficient to have a solution. By definition, one must solve the equations of motion of Type II supergravity.
One can show that the electric Bianchi identities \eqref{eq:elBIS} are implied by \eqref{BPS1}-\eqref{BPS3} when their magnetic cousins \eqref{eq:magBIS} are assumed to hold. Further \eqref{BPS2} implies $V\wedge d_{H_3}f_{\pm}=0$ when $d H_3=0$.  The integrability arguments of \cite{Legramandi:2018qkr} then inform us that, when supersymmetry is preserved, this  amounts to imposing
\beq
dH_3=0,~~~~ \iota_V(d_{H_3} f_{\pm})= 0,~~~~\cos\beta \bigg[d(e^{-2\Phi}\star_8 H_1)+\frac{1}{2}(f_{\pm},f_{\pm})_8\bigg]=0.\label{eq:EOMconds}
\eeq
It then follows that the remaining equations of motion of Type II supergravity are implied, though some additional care is required in the presence of sources for the fluxes, i.e.\ one 
must make sure that they have a supersymmetric embedding and that they come with the appropriate modification of the Bianchi identities.

Let us briefly comment on the Mink$_2$ limit, $m=0$: in general, for such solutions,  $\sin\beta =0$ and the conditions \eqref{eq:noads3cond} are not necessary for supersymmetry to hold (without imposing the first of \eqref{eq:noads3cond}, one necessarily has an additional uncharged U(1) isometry in M$_8$), however one can choose to impose these constraints. When one does,  \eqref{BPS1}-\eqref{BPS6} for $m=0$ provides a set of necessary and sufficient conditions for a subclass of ${\cal N}=(1,1)$ Mink$_2$ solutions; exploring these is outside the scope of this work, but could be interesting.

In the next section we parametrise the bispinors $(\psi_{\pm},\hat{\psi}_{\pm})$ in terms of a $d=8$ SU$(3)$-structure.
\subsection{Parametrising the $d=8$ spinors and G-structures}\label{sec:parameterisation}
We expand the $d=8$ spinors in terms of two unit-norm Majorana--Weyl spinors $\chi_{\pm}$. Such spinors define a G$_2$-structure in eight dimensions, spanned by real 1- and 3-forms $(\tilde{V},\Phi_3)$ in general as
\beq
\chi_{\pm}\otimes \chi_{\pm}^{\dag}=\frac{1}{16}\bigg(1\pm \tilde{V}\wedge\Phi_3-\iota_{\tilde{V}}\star_8 \Phi_3\mp\text{vol}(\text{M}_8) \bigg),~~~~\chi_{\mp}\otimes \chi_{\pm}^{\dag}=\frac{1}{16}\bigg(\tilde{V}\pm \Phi_3-\star_8 \Phi_3\mp\iota_{\tilde{V}}\text{vol}(\text{M}_8)\bigg).
\eeq
We have four Majorana--Weyl spinors that must obey the constraint \eqref{eq: AdS2cond}. An arbitrary $\pm$ chiral spinor may be decomposed in a basis of $(\chi_{\pm},\tilde{U}\chi_{\mp})$, with $\tilde{U}$ a 1-form such that $\iota_{\tilde{V}}\tilde{U}=0$ and $\iota_{\tilde{U}}\tilde{U}=1$. It is possible to show that we can take 
\begin{align}
\chi_{1+}&=\sqrt{c}e^{\frac{1}{2}A}\cos\left(\frac{\beta}{2}\right)\chi_+,~~~~\chi_{1-}=\sqrt{c}e^{\frac{1}{2}A}\sin\left(\frac{\beta}{2}\right)(a\chi_-+ b \tilde{U}\chi_+),\\
\chi_{2-}&=\sqrt{c}e^{\frac{1}{2}A}\sin\left(\frac{\beta_{\pm}}{2}\right)\chi_-,~~~~\chi_{2+}=\sqrt{c}e^{\frac{1}{2}A}\cos\left(\frac{\beta_{\pm}}{2}\right)(a\chi_++ b \tilde{U}\chi_-),
\end{align}
where $a^2+b^2=1$ and
\beq
\beta_+=\beta,~~~~\beta_-=\beta+\pi,
\eeq
without loss of generality. The presence of $\tilde{U}$ means that the G$_2$-structure generically decomposes in terms of an SU$(3)$-structure as follows
\beq
\Phi_3=-(J\wedge \tilde{U}+\text{Re}\Omega),~~~~\iota_{\tilde{V}}\star_8\Phi_3=\frac{1}{2} J\wedge J- \tilde{U}\wedge \text{Im}\Omega,~~~~ J\wedge J\wedge J=\frac{3}{4}i \Omega\wedge \overline{\Omega}.
\eeq
We find it useful to introduce the following complex 1-form and SU$(3)$-structure bilinears on the space orthogonal to this:
\begin{align}
U+i V&:=(a+i b)(\tilde{U}+i \tilde{V}),~~~~\psi^{\text{SU(3)}}_+:=(a+i b) e^{-i J},~~~~\psi^{\text{SU(3)}}_-:=\Omega,\nn\\
\psi^{(7)}_{\pm}&=\psi^{\text{SU(3)}}_{\pm}+i \psi^{\text{SU(3)}}_{\mp}\wedge U.\label{eq:su3structure}
\end{align}
The bispinors then take the form 
\begin{subequations}
\begin{align}
\psi_{\pm}&=\frac{ e^{A}c}{16}\text{Re}\bigg[\psi^{(7)}_{\pm}+\cos\beta\psi^{(7)}_{\mp}\wedge V \bigg],~~~~
\psi_{\mp}=\frac{e^{A} c }{16}\sin\beta V\wedge  \text{Re}\bigg[\psi^{(7)}_{\mp}\bigg], \label{eq:bilinears1}\\
\hat\psi_{\pm}&=\frac{e^A c}{16}\text{Re}\bigg[\psi^{(7)}_{\pm}\wedge V+\cos\beta\psi^{(7)}_{\pm}\bigg],~~~~
\hat\psi_{\mp}=\pm \frac{e^{A}c}{16}\sin\beta\text{Re}\bigg[\psi^{(7)}_{\mp}\bigg]. \label{eq:bilinears2}
\end{align}
\end{subequations}
We note that when $b=0$ these bispinors define a G$_2$-structure, while in the opposite limit, fixing $a=0$, they define an orthogonal SU$(3)$-structure. Generically, the bispinors define a G$_2\times$G$_2$-structure often referred to as an intermediate SU$(3)$-structure\footnote{When a G-structure is a product group one often refers to it via the largest subgroup common to both factors in the product.} which, when $(a,b)$ are point-dependent, can interpolate between these G-structures as one traverses the internal space.

We note that sufficient spinors solving \eqref{eq:noads3conds} and giving rise to \eqref{eq:bilinears1}-\eqref{eq:bilinears2} can be expressed in terms of a pair of unit-norm $d=7$ Majorana spinors $(\chi^{(7)}_1, \chi^{(7)}_2)$ as
\beq
\chi_1= \sqrt{e^A c}\left(\begin{array}{c}\cos\left(\frac{\beta}{2}\right)\\\sin\left(\frac{\beta}{2}\right) \end{array}\right)\otimes \chi^{(7)}_1,~~~~\chi_2= \sqrt{e^A c}\left(\begin{array}{c}\cos\left(\frac{\beta_{\mp}}{2}\right)\\\sin\left(\frac{\beta_{\mp}}{2}\right) \end{array}\right)\otimes \chi^{(7)}_2,
\eeq
where one must decompose the gamma matrices and intertwiner as $\gamma_{a}=\sigma_2\otimes\gamma^{(7)}_{a}$ for $a = 1, \dots, 7$, $\gamma_8= \sigma_1\otimes \mathbb{I},~~B=\mathbb{I}\otimes B^{(7)}$ and take $i\gamma^{(7)}_{1234567}=\mathbb{I}$. We find
\beq
\psi^{(7)}_{+}-i \psi^{(7)}_-= \frac{1}{8} \sum_{n=1}^7\frac{1}{n!}\chi^{(7)\dag}_2\gamma^{(7)}_{a_n\dots a_1}\chi^{(7)}_1 \e^{a_1\dots a_n},
\eeq
and by decomposing 
\beq
\chi_2^{(7)}=a\chi_1^{(7)}-i b U\chi_1^{(7)}
\eeq
we precisely reproduce \eqref{eq:bilinears1}-\eqref{eq:bilinears2} and align $V= \e^8$.

The second condition in \eqref{BPS1} can be locally solved in general by introducing a local coordinate $\rho$ and a function of this coordinate $e^{k}$ such that
\beq
e^{A}\sin\beta V= e^{k} d\rho,\label{eq:localcoord}
\eeq
where $e^{k}$ parametrises diffeomorphisms in $\rho$ so can be set to any convenient non-zero value. The presence of the additional vielbein direction $U$ then means that  one can decompose the internal 8-manifold as
\beq
ds^2(\text{M}_8)=  ds^2(\text{M}_6)+ U^2+ \frac{e^{-2A+2k}}{\sin^2\beta}  d\rho^2,
\eeq
where there exist coordinates with respect to which $U$ and $\text{M}_6$ have no legs/components on $\rho$ but can have functional dependence on it\footnote{i.e.\ $\tilde{k}=e^{A}\sin\beta V$ defines an almost product structure that is integrable by the second of \eqref{BPS1}. This ensures we can choose local coordinates like this.} and M$_6$ is spanned by the vielbein directions that make up $(J,\Omega)$. The types of $\text{M}_6$ that are compatible with supersymmetry will be explored in section \ref{sec:torsionclasses} in terms of SU$(3)$-structure torsion classes.

In the next section we will perform a highly non-trivial check of the conditions we have derived so far by recovering  known classes of AdS$_3$ solution in Type II and $d=11$ supergravities, modulo duality.
\section{\texorpdfstring{Recovering known classes of AdS$_3$ solutions with duality}{Recovering known classes of Ad(3) solutions with duality}}\label{sec:recoveringAdS}
There are several existing G-structure classifications that should be recoverable as certain limits of \eqref{BPS1}-\eqref{BPS6}. Clearly, any AdS$_2$ solution in massless Type IIA supergravity can be lifted to a solution in $d=11$ supergravity. This can result in two types of solutions:  when one has $f_8=0$ the lifted solution will contain a round AdS$_2$ factor, conversely when $f_8\neq 0$ the $d=11$ circle becomes fibered over AdS$_2$, resulting in general in a squashing (and when $f_2\neq 0$ also a fibering over the internal space) of the Hopf fibration of AdS$_3$. Another way to arrive at a supersymmetric AdS$_2$ solution is to take an existing AdS$_3$ solution\footnote{As explained in \cite{Conti:2023naw}, this process only results in round AdS$_2$ when starting from AdS$_3$ solutions with purely magentic NSNS 3-form flux.} and perform T-duality on the Hopf fiber of AdS$_3$ or  SL$(2)$ T-duality on the entire space \cite{Conti:2023naw}. Our system \eqref{BPS1}-\eqref{BPS6} should thus reproduce the necessary geometric conditions for each of these classes of solutions. Establishing  that this is indeed the case provides a highly non-trivial check of our results.

There exists a G-structure classification of supersymmetric AdS$_2$ solutions in $d=11$ in \cite{Hong:2019wyi}.  The case of squashed and fibered AdS$_3$ has not received much attention in the literature, but the simpler case of round AdS$_3$ was considered first in \cite{Martelli:2003ki} and later in  \cite{Tsimpis:2005kj,Legramandi:2020txf}. Finally, supersymmetric AdS$_3$ solutions of Type II supergravity received the G-structure treatment, in the bispinor approach we utilise for AdS$_2$, across \cite{Dibitetto:2018ftj,Passias:2019rga,Passias:2020ubv,Macpherson:2021lbr}. In this section, as a test of what we have derived, we shall recover the geometric conditions defining all known minimally supersymmetric examples of these classes. The Type IIA reduction of \cite{Hong:2019wyi} is likewise recoverable but showing this is a more lengthy computation, so we shall not present it here.

In this section we fix the inverse AdS$_2$ radius $m=1$ for simplicity.

\subsection{\texorpdfstring{AdS$_3$ solutions in $d=11$ supergravity}{AdS(3) solutions in $d=11$ supergravity}}
First, we consider supersymmetic AdS$_3$ solutions of $d=11$ supergravity presented in \cite{Legramandi:2020txf}. In general the map between the bosonic fields of $d=11$ supergravity and massless Type IIA supergravity is
\beq
ds^2_{11}= e^{-\frac{2}{3}\Phi}ds^2+ e^{\frac{4}{3}\Phi} (dz+ C_1)^2,~~~~dC_1= F_2,~~~~ G= F_4+ (dz+C_1)\wedge H,
\eeq  
where $z$ spans the reduction isometry which is assumed to be $2\pi$ periodic and $(F_2,F_4,H)$ are $d=10$ fluxes of the Type IIA theory. For the case at hand, we want to map to a reduction on the Hopf fiber of AdS$_3$, i.e.\
\begin{align}
ds^2_{11}&= \frac{e^{2\Delta}}{4}ds^2(\text{AdS}_2)+ e^{2\Delta}(dz+ \frac{1}{2}\eta)^2+ ds^2(\hat{\text{M}}_8),~~~~ d\eta=\text{vol}(\text{AdS}_2),\nn\\
G&= \frac{1}{4}e^{2\Delta} \text{vol}(\text{AdS}_2)\wedge(dz+ \frac{1}{2}\eta)\wedge G_1 +G.
\end{align}
We must thus constrain the Type IIA fields such that
\beq
f_+=  f_4+f_8,~~~~ H_3=0,~~~~ e^{A}=\frac{1}{2}e^{\Phi}= \frac{1}{2} e^{\frac{3}{2}\Delta},~~~~ ds^2(\text{M}_8)=e^{\Delta}ds^2(\hat{\text{M}}_8),
\eeq
which in $d=10$ (non-democratic) language means we constrain $F_4$ to be purely magnetic, $(F_2,H)$ to be purely electric and set the remaining fluxes to zero. We already know  that the internal space of supersymmetric AdS$_3$ solutions in $d=11$ supports a G$_2$-structure, hence we can fix $(a=1,b=0)$. The conditions \eqref{BPS1}-\eqref{BPS3} and the remaining geometric conditions reduce to 
\begin{subequations}
\begin{align}
& e^{2A}\star_8\lambda f_8 =\frac{1}{2},\label{eq:AdS311d0}\\
&d(e^{3\Delta}\cos\beta)+e^{3\Delta}H_1-2 e^{\frac{3}{2}\Delta} \sin\beta V=0,~~~~d(e^{\frac{3}{2}\Delta}\sin\beta V)=0,\label{eq:AdS311d1}\\
&d(\iota_V\star_8 \Phi_3-\cos\beta V\wedge \Phi_3)= -e^{\frac{3}{2}\Delta}\sin\beta V\wedge f_{4},\label{eq:AdS311d2}\\
&d(e^{\frac{3}{2}\Delta}\sin\beta\Phi_3)-2 (\iota_V\star_8 \Phi_3-\cos\beta V\wedge \Phi_3)-e^{3\Delta}(\star f_4+\cos\beta f_4)=0 ,\label{eq:AdS311d3}
\end{align}
\end{subequations}
the condition \eqref{BPS4s} becomes trivial, while \eqref{BPS5}-\eqref{BPS6} become
\begin{subequations}
\begin{align}
& e^{-3 \Delta}\star_8( 6d\Delta+2\cos\beta H_1)+e^{-\frac{3}{2}\Delta}\sin\beta\Phi_3\wedge  f_4=0,\label{eq:AdS311d4}\\
& e^{-3 \Delta}\star_8(6 e^{-\frac{3}{2}\Delta}  \sin\beta V -6 \cos\beta d\Delta- 2H_1)-e^{-\frac{3}{2}\Delta}\sin\beta\Phi_3\wedge \star_8 f_4=0.\label{eq:AdS311d5}
\end{align}
\end{subequations}
Finally, \eqref{BPS8dpairing} yields
\beq\label{eq:AdS311d7}
\text{vol}(\text{M}_8)+\left(\iota_V\star_8 \Phi_3-\cos\beta V\wedge \Phi_3\right)\wedge f_4= 8 e^{-3\Delta}\left(1-\frac{e^{\frac{3\Delta}{2}}}{4}\sin\beta \iota_V H_1\right)\text{vol}(\text{M}_8).
\eeq
The condition \eqref{eq:AdS311d0} simply gives the $d=10$ RR 2-form we expect,  $F_2=\frac{1}{2}\text{vol}(\text{AdS}_2)$, while the conditions \eqref{eq:AdS311d1}-\eqref{eq:AdS311d5} precisely reproduce the geometric conditions for supersymmetric AdS$_3$ presented in (5.3a)-(5.3f) of \cite{Legramandi:2020txf}. One needs to identify\footnote{Note that the Hodge duals in Type IIA and $d=11$ are taken with respect to different 8-manifolds, and the conventions for the Hodge dual itself (i.e.\ whether the components of a form are taken to contract with the leftmost or rightmost indices of the Levi--Civita symbol) are the opposite.}
\begin{align}
\tilde{f}&=-\cos\beta,~~~~\tilde{\Psi}_3=e^{-\frac{3}{2}\Delta}\sin\beta\Phi_3,~~~~\tilde{\Psi}_4=e^{-2\Delta}(\iota_V\star_8 \Phi_3-\cos\beta V\wedge \Phi_3),~~~~\tilde{F}_1=H_1=G_1\nn\\
\tilde{K}&=e^{-\frac{1}{2}\Delta} \sin\beta V,~~~~\tilde{A}=\Delta,~~~~\tilde{\star}_8 C_p= (-1)^{p}e^{-\frac{\Delta}{2}(8-2p)}\star C_p,~~~~\tilde{F}_4=f_4=G_1,~~~~\tilde{m}=1,
\end{align}
where we add a tilde to the objects appearing in  (5.3a)-(5.3f) of  \cite{Legramandi:2020txf}. The condition \eqref{eq:AdS311d7} is not quoted in \cite{Legramandi:2020txf}, but as \eqref{eq:AdS311d4}-\eqref{eq:AdS311d5} imply this we need not worry. Of course our earlier claim that  \eqref{BPS4s}-\eqref{BPS6} are redundant, which we establish it is indeed true in section \ref{sec:torsionclasses}, implies that the 7-form constraints in \cite{Legramandi:2020txf} are actually implied by the rest of the conditions presented there and an 8-form constraint following from \eqref{eq:AdS311d7}.  

\subsection{\texorpdfstring{AdS$_3$ solutions in Type II supergravity}{AdS(3) solutions in Type II supergravity}}
In general an AdS$_3$ solution in Type II supergravity is decomposable in the form
\begin{align}
\label{eq:1.0}
ds^2&=e^{2A_7}ds^2(\text{AdS}_3)+ ds^2(\text{M}_7),\nn\\
H^{(10)}&=c_0\text{vol}(\text{AdS}_3)+H_7,~~~~ F_7= f_{7\pm}+ e^{3A_7}\text{vol}(\text{AdS}_3)\wedge\star_7 \lambda(f_{7\pm}),
\end{align}
where $(e^{2A_7},f_{7\pm},H_7)$ and the dilaton $\Phi_7$ are defined on $\text{M}_7$, $c_0$ is a constant and the upper/lower signs are taken in Type IIA/IIB. As explained in \cite{Conti:2023naw}, we are free to T-dualise on the Hopf fiber of AdS$_3$ and preserve supersymmetry,\footnote{AdS$_3$ solutions can support two distinct types of supersymmetries of opposite chirality, the duality can preserve all of one chirality and project out those of the opposite chirality. Which chirality is preserved is a choice one is free to make.}as long as we fix
\beq
c_0=0 ~~ \Rightarrow ~~ H^{(10)}=H_7.\label{eq:typeiiads3conds}
\eeq
If we start in Type IIB/IIA we will end up with an AdS$_2$ solution in Type IIA/IIB of the following form\footnote{With respect to \cite{Conti:2023naw} we fix some free signs below for convenience.}
\begin{align}
ds^2&= \frac{e^{2A_7}}{4} ds^2(\text{AdS}_2)+ e^{-2A_7} d\psi^2+ ds^2(\text{M}_7),~~~~e^{- \Phi}= e^{-\Phi_7+A_7},\nn\\
 H&= - \frac{1}{2}d\psi \wedge \text{vol}(\text{AdS}_2)+H_7,~~~~F= f_{7\mp}\wedge d\psi\pm \frac{1}{4}e^{3A_7}\text{vol}(\text{AdS}_2)\wedge \star_7 \lambda (f_{7\mp}). \label{eq:U1Tdualsss}
\end{align}
Comparing the above and the form of $H_1$ in \eqref{BPS1} it is clear that such solutions should lie within our AdS$_2$ classification for 
\begin{align}
e^{2A}&= \frac{1}{4}e^{2A_7},~~~~e^{- \Phi}= e^{-\Phi_7+A_7},~~~~ V= -e^{-A_7}d\psi,~~~~e^{2A}H_1=-\frac{1}{2} d\psi,~~~~H_3=H_7\nn\\
f_{\pm}&=f_{7\mp}\wedge d\psi,~~~~ \star_8 \lambda (f_{\pm})=\pm e^{A_7} \star_7 \lambda (f_{7\pm}),~~~~\sin\beta=1,
\end{align}
where we note that the conditions involving $f_{\pm}$ are indeed consistent, as in our conventions $\star_8 \lambda (f_{7\pm}\wedge V)=\pm\star_7 \lambda (f_{7\pm})$. This makes $(\psi_{\pm},\hat\psi_{\mp})$  strictly orthogonal to $V$ and so, defining 
\beq
\psi_{\pm}= \pm \frac{1}{4} \psi_{7\pm},~~~~ \psi_{\mp}= - \frac{1}{4} \psi_{7\mp},~~~~\text{vol}(\text{M}_7)=\star_8 V,
\eeq
which is consistent as $\cos\beta=0$, the remaining supersymmetry conditions  \eqref{BPS2}-\eqref{BPS8dpairing} reduce to 
\begin{align}
&d_{H_7}(e^{A_7-\Phi_7}\psi_{7\pm})=0,~~~~ d(e^{2A_7-\Phi_7}\psi_{\mp})\pm 2  e^{A_7-\Phi_7}\psi_{7\pm}=\frac{1}{8}ce^{3A_7}\star_7\lambda (f_{7\mp}),\nn\\
&(\psi_{7\pm},f_{7\mp})_7=\pm \frac{1}{2}ce^{-\Phi} \text{vol}(\text{M}_7).
\end{align}
 These conditions precisely reproduce the necessary and sufficient conditions for supersymmetric AdS$_3$ presented in (B.17) of \cite{Macpherson:2021lbr}, in the limit that $c_-=0$ as \eqref{eq:typeiiads3conds} demands, and where the free constant is fixed to $c_+=2 c$. The only not obviously redundant condition that \eqref{BPS4s}-\eqref{BPS6} produce is
\beq
(\psi_{7\mp},f_{7\mp})_6=\pm\frac{1}{4}e^{A_7-\Phi_7}\star_7 dA_7,
\eeq
where  we have made use of the pairing identity $(\psi_{\pm},\star_7\lambda(f_{7\pm}))=\pm (\psi_{\mp}, f_{7\pm})$. But this condition is just another way to write  (B.15) of \cite{Macpherson:2021lbr} which was shown to be redundant there by more involved methods. This is consistent with our earlier claim that  \eqref{BPS4s}-\eqref{BPS6} are redundant.  We have thus established that \eqref{BPS1}-\eqref{BPS8dpairing} are consistent with the Hopf fiber T-dual of AdS$_3$ solutions providing another non-trivial check of our system. 

\section{SU$(3)$-structure torsion classes analysis}\label{sec:torsionclasses}
In this section we shall classify the possible internal spaces and physical fields that supersymmetric AdS$_2$ solutions can support in terms of SU$(3)$-structure torsion classes.

In section \ref{sec:parameterisation} we established that the internal manifold M$_8$ generically supports an SU$(3)$-structure defined on a submanifold M$_6$ orthogonal to the vielbein directions $(U,V)$. There is an enhancement to G$_2$ when $b=0$, but as the forms of a G$_2$-structure can be decomposed in terms of those of the SU$(3)$-structure it is appropriate to perform our general analysis in terms of the torsion classes associated to SU$(3)$. 

In general the torsion classes of a $d=8$ SU$(3)$-structure are a simple extension of the well known $d=7$ case \cite{DallAgata:2003txk}. Indeed if one formally fixes one of the 1-forms $(U,V)$ to zero the torsion classes must reproduce this $d=7$ case; what remains to consider is the portions of $(dJ,d\Omega)$ that have ``legs'' in $U\wedge V$, that of $dV$ with ``legs'' in $U$ and vice-versa. These extra terms should be consistent with
\beq
J\wedge J\wedge J=\frac{3 }{4}i\Omega\wedge \overline{\Omega},~~~~ J\wedge \Omega=0
\eeq
under the exterior derivative, which means it is not hard to show that the $d=8$ torsion class decomposition is
\begin{align}\label{eq:torsions}
dU&=R J+T_1+\text{Re}(\iota_{\overline{{\cal V}}_1}\Omega)+U\wedge W_0+V\wedge U_0+P_0 U\wedge V,\nn\\
dV&=\tilde{R}J+\tilde{T}_1+\text{Re}(\iota_{\overline{\tilde{{\cal V}}}_1}\Omega)+V\wedge \tilde{W}_0+U\wedge \tilde{U}_0+\tilde{P}_0 U\wedge V,\nn\\
dJ&=\frac{3}{2}\text{Im}(\overline{W}_1\Omega)+W_3+W_4\wedge J+U\wedge\bigg(\frac{2}{3}\text{Re}E_1 J+T_2+\text{Re}(\iota_{\overline{{\cal V}}_2}\Omega)\bigg)\nn\\
&+V\wedge\bigg(\frac{2}{3}\text{Re}E_2 J+T_3+\text{Re}(\iota_{\overline{{\cal V}}_3}\Omega)\bigg)+2 \text{Im}{\cal V}_4\wedge U\wedge V,\nn\\
d\Omega&=W_1 J\wedge J+W_2\wedge J+ \overline{W}_5\wedge \Omega+ U\wedge\bigg(E_1\Omega-2 {\cal V}_2\wedge J+S_1\bigg)\nn\\
&+ V\wedge\bigg(E_2\Omega-2 {\cal V}_3\wedge J+S_2\bigg)+ (\iota_{\overline{{\cal V}}_4}\Omega)\wedge U\wedge V,
\end{align} 
On the M$_6$ orthogonal to $(U,V)$ we have: functions $(R,\tilde{R},P_0,\tilde{P}_0)$ that are real and $E_{1,2},W_1$ that are complex, 1-forms $ U_0, \tilde{U}_0,W_0,\tilde{W}_0,W_4$ that are real and  ${\cal V}_1, \tilde{{\cal V}}_1, {\cal V}_2, {\cal V}_3, {\cal V}_4, W_{5}$ that are $(1,0)$-forms, primitive $(1,1)$-forms\footnote{A real $d=6$ $(1,1)$-form, $T$, that is primitive obeys $T\wedge J\wedge J= T\wedge \Omega=0$ by definition.} $T_1,\tilde{T}_1,T_2,T_3$ that are real and  $W_2$ which is complex. Finally, primitive $(2,1)$-forms $S_{1,2}$ which are complex and $W_3$ which is the real part of such a form. These forms transform in irreducible representations of SU$(3)$, and so do not mix with each other. 

$(J, \Omega)$ obey the following useful relations
\begin{align}
\frac{1}{n!}\star_6 J^n&=\frac{1}{(3-n)!}J^{3-n},~~~~\star_6 \Omega= i \Omega,\label{eq:identitiestorsion1}
\end{align}
where in this notation, $J^n=\wedge_{k=1}^n J^k$  for  $n>0$,  $J^0=1$ for $n=0$ and $J^n=0$ for $n<0$. The torsion classes on the other hand obey 
\begin{align}
{\cal V}_i\wedge \Omega&=0,~~~~\frac{1}{n!}\star_6({\cal V}_i\wedge J^n)= i \frac{1}{(2-n)!}{\cal V}_i\wedge J^{2-n},~~~~\star_6 (\overline{{\cal V}}_i\wedge \Omega)=i \iota_{\overline{{\cal V}}_i}\Omega\nn\\
\iota_{\overline{{\cal V}}_i}\Omega\wedge \Omega&=0,~~~~ \overline{\iota_{\overline{{\cal V}}_i}\Omega}\wedge \Omega=-4 {\cal V}_i\wedge J\wedge J,~~~~\iota_{\overline{{\cal V}}_i}\Omega\wedge J=-i \overline{{\cal V}}_i\wedge \Omega \nn\\
T_i\wedge \Omega&= T_i\wedge J\wedge J=0,~~~~  \star_6 (T_i\wedge J^n)= -T_i\wedge J^{n-1},\nn\\
S_i\wedge \Omega&=S_i\wedge \overline{\Omega}= S_i\wedge J=0,~~~~  \star_6 S_i= -i S_i\label{eq:identitiestorsion2}.
\end{align}

For the case at hand, given that we can locally take
\beq
V= \frac{e^k}{e^A\sin\beta}d\rho,
\eeq
we can immediately refine  \eqref{eq:torsions}. In order to do this it is useful to decompose the exterior derivative as
\beq
d= \tilde{d}^{(6)}+U\wedge \iota_U d+d\rho\wedge \partial_{\rho}, \label{eq:decompostionofd}
\eeq
where $\tilde{d}^{(6)}$ is a twisted exterior derivative in general.\footnote{ i.e.\ one can introduce a local coordinate $y$ such that $
U=  e^{C}(dy+  {\cal A})$, where ${\cal A}$ has no legs in $(y,\rho)$. Then $
\tilde{d}^{(6)}=  d^{(6)}- {\cal A}\wedge \partial_y$, where $d^{(6)}$ is the usual $d=6$ exterior derivative.} We then have
\beq
\tilde{R}=\tilde{T}_1=\tilde{{\cal V}}_1=\tilde{U}_0=0,~~~~\tilde{W}_0= d^{(6)}\left(\log(e^{A}\sin\beta)\right),~~~~\tilde{P}_0=\iota_U d\left(\frac{1}{e^{A}\sin\beta}\right).
\eeq

When one attempts to solve the conditions \eqref{BPS1}-\eqref{BPS6}, it is possible for several of the torsion classes $W_{1,2,3,4,5}$ to become fixed. This gives us information about the type of manifold that can live on M$_6$, several well-known examples are given in table \ref{table:one}.
\begin{table}[h!]
\centering
\begin{tabular}{||c | c ||} 
 \hline
M$_6$ & Vanishing torsion classes\\ [0.5ex] 
 \hline\hline
Complex & $W_1=W_2=0$\\
Symplectic& $W_1=W_3=W_4=0$\\
Half-flat& $\text{Re}W_1=\text{Re}W_2=W_4=W_5=0$\\
Special Hermitian& $W_1=W_2=W_4=W_5=0$\\
Nearly K\"ahler& $W_2=W_3=W_4=W_5=0$\\
Almost K\"ahler& $W_1=W_3=W_4=W_5=0$\\
K\"ahler&  $W_1=W_2=W_3=W_4=0$\\
Calabi--Yau&  $W_1=W_2=W_3=W_4=W_5=0$\\
 \hline
\end{tabular}
\caption{\small A table of well-known SU$(3)$-structure manifolds in terms of vanishing torsion classes. Note that a manifold may be conformally one of these for certain values of $(W_4,W_5)$.}
\label{table:one}
\end{table}

To proceed,  we further decompose the fluxes as
\begin{align}
f_{\pm}&=  g^{1}_{\pm}+ U \wedge g_{\mp}^1 +  V\wedge g_{\mp}^2 + U \wedge V\wedge g^{2}_{\pm},\nn\\
H_3&=  H^{(6)}_3+ U \wedge H^1_2 +  V\wedge H^2_2 + U \wedge V\wedge H_1^{(6)},
\end{align}
and then introduce a $d=6$ Hodge dual such that the RR flux, or indeed any form in $d=8$, behaves as
\begin{align}
\star \lambda (f_{\pm})&= -\star_6 \lambda(g^2_{\pm})+ U\wedge \star_6 \lambda(g^2_{\mp})-V\wedge \star_6 \lambda(g^1_{\mp})+U\wedge V\wedge \star_6 \lambda(g^1_{\pm}),\nn\\
\star f_{\pm}&= \star_6 g^2_{\pm}\pm U\wedge \star_6 g^2_{\mp}\mp V\wedge \star_6 g^1_{\mp}+U\wedge V\wedge \star_6 g^1_{\pm}.
\end{align}
Note we also have
\beq
\star_6\star_6 C_k=(-)^k C_k,~~~~\lambda (\star_6 C_k)=-(-)^k\star_6 \lambda (C_k).
\eeq
Examining \eqref{BPS1}-\eqref{BPS8dpairing}, at first sight, it appears that these merely serve to fix the RR flux in terms of SU$(3)$-structure torsion classes and the NSNS 3-form. However, one should appreciate that, as\footnote{If this were not the case, and one did fix $\cos\beta=1$, \eqref{BPS3} would contain no information about the anti-self-dual 4-form components of $f_+$ in Type IIA.} $\cos\beta \neq 1$,  \eqref{BPS3} can be manipulated to simply define the entire RR flux on its own as
\beq
e^{2A} \sin^2\beta c f_{\pm}=  \pm 16 \left(\cos\beta- \star_8 \lambda \right)\left(d_{H_3}(e^{A-\Phi}\hat\psi_{\pm})-m e^{-\Phi} \psi_{\pm}\right).\label{eq:flux def}
\eeq
Given this, the remaining conditions then serve to fix the NSNS flux and the possible form of M$_8$ through restrictions to the SU$(3)$-structure torsion classes and physical fields. The content of \eqref{BPS1} merely defines the electric part of the NSNS flux and defines the local coordinate $\rho$  in \eqref{eq:localcoord}.  We introduce the shorthand notation
\beq
d^{(7)}:= \tilde{d}^{(6)}+ U\wedge \iota_U d,~~~~ f_{\pm}^{(7)}:= g_{\pm}^1+ U\wedge g^1_{\mp}.
\eeq
The content of \eqref{BPS2} is simply \eqref{BPS:decendent}, which already only involves the NSNS flux, and a second equation for the part of the RR flux orthogonal to $d\rho$
\begin{align}
f_{\pm}^{(7)}&=\left(d^{(7)}-(H^{(6)}_3+U\wedge H^1_2)\wedge \right)\left(e^{-\Phi}\cot\beta\text{Re}\psi^{(7)}_{\mp}\right)\pm e^{-\Phi}\csc\beta (H^2_2+U\wedge H^{(6)}_1)\wedge \text{Re}\psi^{(7)}_{\pm}\nn\\
&\mp e^{-k}\partial_{\rho}\left(e^{A-\Phi}\text{Re}\psi^{(7)}_{\pm}\right).
\end{align}
Making this consistent with \eqref{eq:flux def} leads to further restrictions on the NSNS sector. Finally, one can use the previous expressions to eliminate $(H_1,f_{\pm})$ from  \eqref{BPS8dpairing} and to extract further restrictions on the geometry and physical fields.

In the next sections we present all the conditions that the above considerations imply; to derive these we make copious use of the identities in  \eqref{eq:identitiestorsion1}-\eqref{eq:identitiestorsion2}. Needless to say, this is a long and tedious computation, so we omit the details.

\subsection{Type IIA}
In this section we present the results of our torsion classes analysis in Type IIA supergravity.

Considering first only the differential constants \eqref{BPS1}-\eqref{BPS3} we find the following conditions on the fields and functions of the spinor Ansatz
\beq
\tilde{d}^{(6)}(e^{A-\Phi}a)=\iota_U d(e^{A-\Phi}a)=0,
\eeq
which imply $e^{A-\Phi}a$ is a function of $\rho$ in general. The magnetic components of the NSNS flux get fixed in terms of the torsion classes as
\begin{align}
H^{(6)}_1&=4\text{Im}(b {\cal V}_3- a{\cal U}_1),\nn\\
H^1_2&=-b\text{Im}W_2+\frac{2}{3}a e^{-A+\Phi}\iota_Ud(e^{A-\Phi}b) J+\text{Re}\iota_{\overline{\cal U}_2}\Omega,\nn\\
H^2_2&= H^{(1,1)}+\text{Re}\iota_{\overline{\cal U}_1}\Omega+ \frac{2}{3}a^2\partial_{\rho}\left(\frac{b}{a}\right) e^{A-k}\sin\beta J,\nn\\
H^{(6)}_3&= \text{Re} H^{(2,1)}+\lambda\text{Im}\Omega-\frac{1}{2}e^{-A+\Phi}\iota_{U}d(e^{A-\Phi} b)\text{Re}\Omega+\left(e^{-A+\Phi} a \tilde{d}^{(6)}(e^{A-\Phi}b)+2 b\text{Im}{\cal V}_1\right)\wedge J.
\end{align}
where we introduce the function $\lambda$, (1,0)-forms ${\cal U}_i$ and primitive $(1,1)$ and $(2,1)$ forms $(H^{(1,1)},H^{(2,1)})$.  We find that the following torsion classes  are fixed in general
\begin{align}
T_2&=-a \text{Im}W_2,~~~~\text{Re}{\cal V}_2=a\left[\frac{1}{2}\left(W_0-\tilde{d}^{(6)}(A-\Phi)\right)-\text{Re}W_5\right],\nn\\
\text{Im}{\cal V}_4&=-\frac{1}{2}\left( U_0^{D}-4\text{Im}(b{\cal U}_1+a{\cal V}_3)\right)+\frac{1}{2}e^{2A}\sin^2\beta \tilde{d}^{(6)}\left(\frac{\cos\beta}{e^{2A}\sin^2\beta}\right),\nn\\
W_4&=-b e^{-A+\Phi}\tilde{d}^{(6)}(e^{A-\Phi}b)+2a \text{Im}{\cal V}_1,~~~~\text{Im}E_2=-\frac{1}{2}e^{2A}\sin^2\beta\iota_U d\left(\frac{\cos\beta}{e^{2A}\sin^2\beta}\right),\\
\text{Re}E_2&=\frac{1}{2}(P_0-2m e^{-A}\cot\beta)-\frac{1}{2}e^{-2A+2\Phi-k}\partial_{\rho}(e^{3A-2\Phi}\sin\beta),\nn\\
\text{Re}E_1&=-\frac{3}{2}\left(a\text{Im}W_1+b e^{-A+\Phi}\iota_Ud(e^{A-\Phi}b)\right),\nn
,\nn
\end{align}
where a superscript $D$ is such that for a complex $(p,q)$-form $\omega$, $(\text{Re}\omega)^D = \text{Im}\omega$; it follows that for a real 1-form $\alpha$, $\alpha+i\alpha^D$ is a (1,0)-form. We find additional constraints on the torsion classes, whose solution requires one to make assumptions about the values of $(a,b)$, and thus likely define a branching of possible classes
\begin{align}
b W_1&=\frac{a}{3}\left(2\lambda+i e^{-A+\Phi}\iota_{U}d(e^{A-\Phi}b)\right),~~~~ b W_3=a \text{Re} H^{(2,1)}.
\end{align}
Moving onto the pairing constraint \eqref{BPS8dpairing} we find it gives rise to the differential condition 
%\vspace{-2mm}
\beq
\begin{gathered}
\left[2 e^{A}(2\text{Im}E_1+3 R+6 a \text{Re}W_1+4 b \lambda)\sin\beta+m(7+\cos(2\beta))\right]e^{k}=2 e^{4A}\sin^4\beta\partial_{\rho}\left(\frac{\cos\beta}{e^{2A}\sin^2\beta}\right).
\end{gathered}
\eeq
This is all that is contained in \eqref{BPS1}-\eqref{BPS8dpairing}, that does not serve to fix the form of the RR fluxes. These are as follows:

The components of $g^1_{\pm}$ are 
\begin{align}
e^kg^1_0&= \partial_{\rho}(e^{A-\Phi}a),~~~~e^{2A}\sin\beta g_1^1=e^{2A-\Phi}b (U^D_0- W_0\cos\beta)- 4 e^{2A-\Phi} U^D_0+\cot\beta \tilde{d}^{(6)}(e^{2A-\Phi} b \sin\beta),\nn\\[2mm]
e^{2A+k}\sin\beta g^1_2&= -e^{2A+k-\Phi}\left[a(H^{(1,1)}+\text{Re}\iota_{\overline{\cal U}_1}\Omega)+ b\cos\beta(T_1+\text{Re}\iota_{\overline{\cal V}_1}\Omega)-b(T_3+\text{Re}\iota_{\overline{\cal V}_3}\Omega)\right]\nn\\
&-\frac{1}{3}\left[e^{A-\Phi+k} b(2m \cot\beta+3 e^{A}\cos\beta R-e^A P_0)- e^{3A}\sin^2\beta \partial_{\rho}\left(\frac{e^{-\Phi} b}{\sin\beta}\right)\right]J,\nn\\[2mm]
e^{2A+k}\sin\beta g^1_3&= e^{2A+k-\Phi}\left(\text{Im}S_2-\cos\beta(b \text{Re} H^{(2,1)}+ a W_3-\text{Re}S_1)\right)+\frac{1}{2}e^{2A+k-\Phi}\iota_{U}d(\cos\beta)\text{Re}\Omega\nn\\
&-\frac{1}{2}e^{A-\Phi}\left[e^A\partial_{\rho}(e^A\sin\beta)+ e^k\left(e^AP_0+e^A(2\text{Im}E_1+3 a\text{Re}W_1+2 b \lambda)\cos\beta+2m \cot\beta\right)\right]\text{Im}\Omega\nn\\
&+e^{2A+k-\Phi}\left[2\cos\beta(a\text{Re}W_5-\text{Im}{\cal V}_1)+2 \text{Im}{\cal V}_3-a U^D_0-a\cos\beta \tilde{d}^{(6)}\log\left(\frac{e^{\Phi}}{\sin\beta}\right)\right]\wedge J,\nn\\[2mm]
e^{2A+k}\sin\beta g^1_4&=e^{2A+k}\left[\sin\beta \tilde{d}^{(6)}(e^{-\Phi}\cot\beta)+ e^{-\Phi}\left((-1+2b)U_0^D+2\cos\beta(\text{Re}W_5-a \text{Im}{\cal V}_1)+2 a\text{Im}{\cal V}_3\right)\right]\wedge \text{Re}\Omega\nn\\
&+ e^{2A+k-\Phi}\left(\cos\beta(\text{Re}W_2+a T_1)- b H^{(1,1)}-a T_3\right)\wedge J\nn\\
&+\frac{1}{6}\left[\csc\beta \partial_{\rho}\left(e^{3A-\Phi}a\sin^2\beta\right)+2 e^{A+k-\Phi}\left(-a e^{A} P_0+3 e^A(a R+\text{Re}W_1)\cos\beta+ 2 a m \cot\beta\right)\right]J^2,\nn\\[2mm]
\sin\beta g^1_5&=\frac{1}{2}\left[\sin\beta \cos\beta \tilde{d}^{(6)}\left(\frac{e^{-\Phi}b}{\sin\beta}\right)-e^{-\Phi}\left(b U_0^D+\cos\beta(-b W_0+4 \text{Re}{\cal U}_2)\right)\right]\wedge J^2,\\
e^{A+k}\sin\beta g^1_6&=\frac{1}{3!}\left[e^{k -\Phi}\left(-e^{A}b P_0+ e^{A}(3 b R+ 4 \lambda)\cos\beta+2 bm \cot\beta\right)+ \partial_{\rho}(e^{2A-\Phi} b\sin\beta)\right] J^3,\nn
\end{align}
and for  $g^2_{\pm}$ we find
\begin{align}
e^{2A+k}\sin\beta g^2_0&= e^A\cos\beta\partial_{\rho}\left(e^{2A-\Phi} b \sin\beta\right)\nn
&+ e^{A+k-\Phi}\left(b(m\sin\beta- e^{A}\cos\beta P_0+2 m \cos\beta \cot\beta )+e^{A}(3b R+4 \lambda)\right),\nn\\[2mm]
e^{2A}\sin\beta g^2_1&= -e^{2\Phi}\sin\beta\left(\tilde{d}^{(6)}\left(\frac{e^{2A-3\Phi}b}{\sin\beta}\right)\right)^D+ e^{2A-\Phi}b(W_0^D-4\text{Im}W_5- \cos\beta U_0),\nn\\[2mm]
e^{2A+k}\sin\beta g^2_2&= e^{2A+k-\Phi}\left(\text{Re}W_2+ a T_1-\cos\beta( b H^{(1,1)}+a T_3)\right)\nn\\
&+\frac{1}{3}e^{2A}\left(a e^{k-\Phi}(2 \text{Im}E_1-3 R+2 \cos\beta P_0)- \sin\beta \partial_{\rho}(e^{A-\Phi}a \cos\beta)\right)J\nn\\
&+e^{k}\left[\frac{1}{2}\text{Im}\iota_{\overline{{\cal U}}_3}\Omega+ e^{2A-\Phi}\left(\text{Im}\iota_{\overline{W}_5}\Omega+\text{Re}\iota_{\left(-a \overline{{\cal V}}_1+\cos\beta(\overline{{\cal V}}_4-a \overline{{\cal V}}_3-b\overline{{\cal U}}_1)\right)}\Omega\right)\right],\nn\\[2mm]
e^{2A+k}\sin\beta g^2_3&=e^{2A+k-\Phi}\left(b \text{Im}H^{(2,1)}+a W^D_3-\text{Im}S_1+\cos\beta \text{Re}S_2\right)+\frac{1}{2}e^{-\Phi+k}\sin\beta \iota_Ud(e^{2A}\sin\beta)\text{Im}\Omega\nn\\
&+e^{A+k-\Phi}\left[e^A\left(2(a\text{Im}W_5+\text{Re}{\cal V}_1)+\cos\beta(a U_0-2 \text{Re}{\cal V}_3)\right)- \frac{e^{A-\Phi} a}{\sin\beta} \left(\tilde{d}^{(6)}(e^{\Phi}\sin\beta)\right)^D\right]\wedge J\nn\\
&+\frac{1}{2}e^{A-\Phi}\left[e^{A}\cos\beta \partial_{\rho}(e^A\sin\beta)+e^k\left(e^{A}\cos\beta P_0+ e^{A}(2\text{Im}E_1+3 a \text{Re}W_1+2 b \lambda)+2 m \csc\beta\right)\right]\text{Re}\Omega,\nn\\[2mm]
e^{2A+k}\sin\beta g^2_4&=e^{2A+k-\Phi}\left( b T_1+\cos\beta(a H^{(1,1)}-b T_3)\right)\wedge J\nn\\
&+2 e^{2A+k-\Phi}\left(b \text{Im}{\cal V}_1+ \cos\beta(a U_0^D-b \text{Im}{\cal V}_3)\right)\wedge \text{Re}\Omega\nn\\
&+\frac{1}{12}\left[e^{A+k-\Phi}b\left(e^A(-6 R+2 P_0 \cos\beta)+m\frac{(-5+\cos(2\beta))}{\sin\beta}\right)+e^{3A}\sin(2\beta)\sin\beta \partial_{\rho}\left(\frac{e^{-\Phi}b}{\sin\beta}\right)\right]J^2,\nn\\[2mm]
e^{2A}\sin\beta g^2_5&=\frac{1}{2}\left[e^{2A-\Phi}\left(b W_0^D+\cos\beta(b U_0-4 \text{Re}{\cal U}_1)\right)-\csc\beta \left(\tilde{d}^{(6)}(e^{2A-\Phi} b \sin\beta)\right)^D\right]\wedge J^2,\nn\\
e^{2A+k} g^2_6&=-\frac{1}{6}\left(e^{2A}\cos\beta \partial_{\rho}(e^{A-\Phi} a)- e^{A+k-\Phi} a m \right) J^3,
\end{align}
where we have introduced
\beq
\text{Re}{\cal U}_3:=\frac{\tilde{d}^{(6)} (e^{2A-\Phi}\sin\beta)}{\sin\beta},
\eeq
to ease the presentation a little.

Given all the above one finds that \eqref{BPS4s}-\eqref{BPS6} are indeed implied as claimed earlier.

\subsection{Type IIB}
In this section we present the results of our torsion classes analysis for Type IIB supergravity.

Considering again first \eqref{BPS1}-\eqref{BPS3}, this time we find a single constraint involving only the functions of our Ansatz
\beq
\tilde{d}^{(6)}(e^{2A-\Phi}a \sin\beta)=0.
\eeq
The magnetic contribution to the NSNS 3-form must decompose in terms of the torsion classes, two functions $\lambda_{1,2}$ and (1,0)-forms ${\cal U}_i$ as
\begin{align}
H^{(6)}_1&=e^{2A}\sin^2\beta \left(\tilde{d}^{(6)}\left(\frac{\cos\beta}{e^{2A}\sin^2\beta}\right)\right)^D+4\text{Im}(b {\cal V}_3-a {\cal U}_2),\\
H^1_2&= H^{(1,1)}_1+ \lambda_1 J +\text{Re}\iota_{\overline{\cal U}_1}\Omega,\nn\\
H^2_2&= H^{(1,1)}_2+ \frac{1}{3}e^{A}\sin\beta\left[e^A\sin\beta \iota_{U}d\left(\frac{\cos\beta}{e^{2A}\sin^2\beta}\right)+2 e^{-k} a^2\partial_{\rho}\left(\frac{b}{a}\right)\right] J+\text{Re}\iota_{\overline{\cal U}_2}\Omega,\nn\\
H^{(6)}_3&= \text{Re} H^{(2,1)}+\lambda_2\text{Re}\Omega+\left[\frac{a}{e^{2A-\Phi}\sin\beta}\tilde{d}^{(6)}\left(e^{2A-\Phi} b \sin\beta\right)+ 2 \text{Re}(a {\cal U}_1- b {\cal V}_2)\right]\wedge J,\nn
\end{align}
while the following torsion classes get fixed in general
\begin{align}
E_1&= -\frac{3}{2}a\text{Im}W_1-\iota_{U}d(A-\Phi) + b\lambda_2 + \frac{3}{2}i  R,~~~~E_2=\frac{1}{2}P_0- m e^{-A}\cot\beta- \frac{1}{2}e^{-2A-k+2\Phi}\partial_{\rho}\left(e^{3A-2\Phi}\sin\beta \right),\nn\\
\text{Re}W_1&=-a R,~~~~ \text{Re}W_2=-a T_1,~~~~\text{Re}W_5=-\frac{1}{2}\tilde{d}^{(6)}(A-\Phi)+a \text{Im}{\cal V}_1,\nn\\
W_4&= a^2\left(W_0+ \tilde{d}^{(6)}\log(e^A\sin\beta)\right)-2 \text{Re}(b {\cal U}_1+a {\cal V}_2),~~~~\text{Re}S_1=a W_3+ b \text{Re} H^{(2,1)},\nn\\
\text{Re}{\cal V}_4&=-\frac{1}{2}U_0+2\text{Re}(a {\cal V}_3+b {\cal U}_2).
\end{align}
We find the following conditions that define a branching of possible classes of solutions
\begin{align}
e^{A-\Phi}b W_0&= \tilde{d}^{(6)}(e^{A-\Phi}b),~~~~ b R=0,~~~~ b T_1=0,~~~~ b {\cal V}_1=0,\nn\\
0&=\cos\beta \left(e^{A}\sin\beta \partial_{\rho}(e^{2A-\Phi}a \sin\beta)+(m e^{A-\Phi+k}a \cos\beta -1)\right),
\end{align}
where we note in particular that the form of $U$ becomes highly constrained when $b\neq 0$, i.e. when we are not in the G$_2$-structure limit.

We find that the pairing constraint \eqref{BPS8dpairing} gives rise to the following differential equation involving the torsion classes.
\beq
2 e^{4A}\sin^3\beta \partial_{\rho}\left(\frac{\cos\beta}{e^{2A}\sin^2\beta}\right)= e^{k}\left[-2 e^A\left(6 b \text{Im}W_1-2 a^2\iota_{U}d\left(\frac{b}{a}\right)+3 \lambda_1+4 a \lambda_2\right)+m\csc\beta\left(7+\cos(2\beta)\right)\right].
\eeq
This just leaves the RR flux to be presented: the components of $g^1_{\pm}$ are the following
\begin{align}
e^{2A+k}g^1_0&= e^{2A}\partial_{\rho}(e^{A-\Phi}b)-b e^{2A+k-\Phi}P_0 \csc\beta+ e^{2A+k}\iota_Ud\left(e^{-\Phi}a\cot\beta \right),\nn\\[2mm]
e^{2A}\sin\beta g^1_1&=e^{2A-\Phi}b U_0+ e^{2A}\sin\beta \tilde{d}^{(6)}\left(e^{-\Phi} a \cot\beta \right),\nn\\[2mm]
e^{2A+k}\sin\beta g^1_2&=- e^{2A+k-\Phi}\left(b H^{(1,1)}_2+\cos\beta(a H_1^{(1,1)}-b T_2)+a T_3\right)-\frac{1}{2} e^{2A+k-\Phi}\text{Re}\iota_{\overline{{\cal U}}_4}\Omega\nn\\
&+\frac{1}{3}\bigg[ \frac{e^{2A+k}\tan\beta}{b^2}\iota_Ud\left(e^{-\Phi}b^3\cos\beta \cot\beta\right)+ e^{2A+k-\Phi}\cos\beta\left(-3a(b\text{Im}W_1+ \lambda_1)+2 b^2\lambda_2\right)\nn\\
&+2 e^{A+k-\Phi}(a e^{A}P_0+ a m \cot\beta)-e^{3A}\sin^2\beta\partial_{\rho}\left(e^{-\Phi}a\csc\beta\right)\bigg] J, \nn\\[2mm]
e^{2A+k}\sin\beta g^1_3&=e^{2A+k-\Phi}\left(\cos\beta(b W_3-a \text{Re} H^{(2,1)})- \text{Re}S_2\right)\nn\\
&+\left[-e^{2A+k}\sin\beta \tilde{d}^{(6)}(e^{-\Phi}b \cot\beta)+ e^{2A+k-\Phi}\left(aU_0+2\text{Re}(\cos\beta{\cal U}_1- {\cal V}_3)\right)\right]\wedge J\nn\\
&+\frac{1}{2}e^{A-\Phi}\left[e^{A}\partial_{\rho}(e^A \sin\beta)- e^{k}\left(e^A P_0+ e^{A}\cos\beta (3 b\text{Im}W_1+2 a \lambda_2)-2m \cot\beta\right)\right]\text{Re}\Omega,\nn\\[2mm]
e^{2A+k}\sin\beta g^1_4&=e^{2A+k-\Phi}\left(a H^{(1,1)}_2-\cos\beta(b H^{(1,1)}_1+\text{Im}W_2+a T_2)-b T_3\right)\wedge J\nn\\
+e^{A+k-\Phi}\big\{-&\cot\beta \tilde{d}^{(6)}(e^{A}\sin\beta)-2e^{A}\text{Re}(a {\cal U}_2-b {\cal V}_3)+ e^{A}\cos\beta\left[W_0-2\text{Re}\left(b {\cal U}_1+a({\cal V}_2- i{\cal V}_1)\right)\right]\big\}\wedge\text{Im} \Omega\nn\\
&+\frac{1}{6}\bigg\{\csc\beta\partial_{\rho}(e^{3A-\Phi} b \sin^2\beta)-\frac{1}{2}e^{A+k-\Phi}\bigg[\frac{e^{A-\Phi}}{a^2 \sin\beta}\iota_U d\left(e^{-\Phi} a^3 \sin(2\beta)\right)\nn\\[2mm]
&+2 b\left(-e^AP_0+2 e^A\cos\beta(3b \text{Im}W_1+3 \lambda_1+2 a \lambda_2)\right)-4m \cot\beta\bigg]\bigg\}J^2,\nn\\
e^{2A+k}\sin\beta g^1_5&=-\frac{1}{2} e^{2A+k-\Phi}\left[a\tilde{d}^{(6)}(\cos\beta)+b U_0-4 \text{Re}{\cal U}_2+2 \cos\beta\left(a W_0-2(\text{Im}{\cal V}_1+ \text{Re}{\cal V}_2)\right)\right]\wedge J^2,\nn\\[2mm]
e^{2A+k}\sin\beta g^1_6&=\frac{1}{6}\bigg\{-e^{A}\partial_{\rho}(e^{2A-\Phi}a \sin\beta)-\cot\beta e^{A+k-\Phi}\bigg[e^{A-\Phi}\iota_Ud(e^{\Phi}b \csc\beta)+ \nn\\
&2a m -e^{A}\sin\beta\left(3a(b \text{Im}W_1+ \lambda_1)+2(1+a^2) \lambda_2\right)\bigg]\bigg\}J^3,
\end{align}
where we have defined
\beq
\text{Re}{\cal U}_4:=2 \cos\beta(a \text{Re}{\cal U}_1-b \text{Re}{\cal V}_2)-2(b\text{Re}{\cal U}_2+a \text{Re}{\cal V}_3)+ U_0.
\eeq
The components of $g^2_{\pm}$ are
\begin{align}
e^{2A+k}\sin\beta g^2_0&=e^{A}\cos\beta \partial_{\rho}(e^{2A-\Phi}a \sin\beta)+e^{A+k-\Phi}\bigg\{\csc\beta e^{A-\Phi}\iota_U d(e^{\Phi}b\sin\beta)\nn\\
&+\frac{1}{2}\left[-6 a e^{A}(b \text{Im}W_1+\lambda_1)- 4 e^A \lambda_2(1+a^2)+am\csc\beta \left(3+\cos(2\beta)\right)\right]\bigg\},\nn\\
e^{A+\Phi}\sin\beta g^2_1&=-e^{-A}\sin\beta a\left( \tilde{d}^{(6)}(e^{2A}\sin\beta)\right)^{D}-e^A\left(2 a W_0^D+ b \cos\beta U_0^D-4(\cos\beta \text{Im}{\cal U}_2+\text{Im}{\cal V}_2-\text{Re}{\cal V}_1)\right),\nn\\[2mm]
e^{2A+k}\sin\beta g^2_2&=e^{2A+k-\Phi}\left(b H_1^{(1,1)}+ \text{Im}W_2+ a T_2+ \cos\beta(b T_3- H^{(1,1)}_2)\right)\nn\\
&+\frac{1}{12}\bigg\{4 \cot\beta \partial_{\rho}(e^{3A-\Phi}b\sin^2\beta) -\frac{4e^{A+k-2\Phi}}{a^2\sin\beta}\iota_Ud(e^{A+\Phi}a^3 \sin\beta)+ e^{-2A+k-\Phi}a \iota_U d\left(e^{4A}\cos(2\beta)\right)\nn\\
&+2 e^{A+k-\Phi}\left[-4 b e^{A}\left(3 (b \text{Im}W_1+ \lambda_1)+2 a \lambda_2\right)+2 b e^{A} P_0 \cos\beta+ b m\csc\beta\left(7+ \cos(2\beta)\right)\right]\bigg\}J\nn\\
&+ e^{2A+k -\Phi}\bigg(a \text{Im}\iota_{\overline{{\cal V}}_1}\Omega+ \text{Re}\iota_{\overline{{\cal U}}_5}\Omega\bigg),\nn\\[2mm]
e^{2A+k}\sin\beta g^2_3&=- e^{2A+k-\Phi}\left( a \text{Im}H^{(2,1)}- b W_3^D+ \cos\beta  \text{Im}S_2\right)\nn\\
&-\frac{1}{2}e^{A-\Phi}\left[e^{A}\cos\beta \partial_{\rho}(e^A \sin\beta)-e^k\left(e^A\cos\beta P_0+3 e^A b \text{Im}W_1+2 e^{A}a\lambda_2-2 m \csc\beta \right)\right]\text{Im}\Omega\nn\\
&-\frac{1}{2}e^k\bigg[e^{A}\left(\tilde{d}^{(6)}(e^{A-\Phi}b)\right)^{D}+\frac{1}{2}e^{6A-\Phi} \sin^4\beta b\left(\tilde{d}^{(6)}\left(\frac{\cos(2\beta)}{e^{4A}\sin^4\beta}\right)\right)^{D}\nn\\
&+ e^{2A-\Phi}\left(b W^D_0-2 a U^D_0\cos\beta+4\text{Im}(\cos\beta {\cal V}_3-{\cal U}_1)\right)\bigg]\wedge J,\nn\\[2mm]
e^{2A+k}\sin\beta g^2_4&=-e^{2A+k-\Phi}\left(a H^{(1,1)}_1-b T_2+ \cos\beta(b H^{(1,1)}_2+ a T_3)\right)\wedge J\nn\\
&+ e^{2A+k -\Phi}\text{Im}\left(2(-a {\cal U}_1+b {\cal V}_2)+\cos\beta(b {\cal U}_2+ a {\cal V}_3- i U_0^D)\right)\wedge\text{Re}\Omega\nn\\
&+\frac{1}{12}\sin\beta \bigg\{ e^{3A}\sin(2\beta)\partial_{\rho}(e^{-\Phi}a \csc\beta)+ e^k\csc\beta\bigg[-\frac{2}{b^2}\iota_Ud(e^{2A-\Phi} b^3)\nn\\
&+ 2e^{-\Phi}b\cos(2\beta)\csc\beta \iota_U d(e^{2A}\sin\beta)+6 e^{2A-\Phi}a(b \text{Im}W_1+\lambda_1)-4 e^{2A-\Phi} b^2\lambda_2\nn\\
&-4 e^{2A-\Phi}a P_0\cos\beta+ e^{A-\Phi}am\csc\beta\left(\cos(2\beta)-5\right)\bigg]\bigg\}J^2,\nn\\[2mm]
e^{A+\Phi}\sin\beta g^2_5&=\frac{1}{2}\cos\beta \left(e^A b U_0+ e^{3A}a \sin^2\beta \tilde{d}^{(6)}\left(e^{-2A}\cot\beta\csc\beta \right)\right)^D\wedge J^2,\\
e^{2A+k}\sin\beta g^2_6&=\frac{1}{6}\bigg[e^{2A}\sin\beta\cos\beta \partial_{\rho}\left(e^{A-\Phi}b\right)+ e^{k}\bigg(e^{A}\iota_Ud(e^{A-\Phi}a)\nn\\
&+\frac{1}{4}e^{6A-\Phi}a \sin^4\beta \iota_U d\left(\frac{\cos(2\beta)}{e^{4A}\sin^4\beta}\right)- e^{A-\Phi}b(\cos\beta e^A P_0+ m \sin\beta)\bigg)\bigg]J^3\nn,
\end{align}
where we have defined
\begin{align}
\text{Re}{\cal U}_5&:=-b \text{Re}{\cal V}_1-a(\cos\beta \text{Re}{\cal U}_2+ \text{Re}{\cal V}_2)+b\cos\beta \text{Re}{\cal V}_3+\frac{1}{2}W_0-\frac{1}{2 }\tilde{d}^{(6)}\log(e^A\sin\beta).
\end{align}
Again, as in Type IIA, the conditions \eqref{BPS4s}-\eqref{BPS6} are implied by the above.

\section{\texorpdfstring{New classes of AdS$_2$ solutions}{New classes of AdS(2) solutions}}\label{eq:examples}
In this section we show the utility of our results by deriving two new interesting classes of AdS$_2$ solutions, one in each of Type IIA and IIB supergravity.

In section \ref{sec:IIAexample} we derive a class of ${\cal N}=1$ solutions in (massive) Type IIA supergravity for which M$_8$ decomposes as a foliation of a weak G$_2$-manifold  over an interval and which includes as a special case the ${\cal N}=8$ class of \cite{Dibitetto:2018gbk}. In section \ref{sec:IIBexample} we derive a broad class of small ${\cal N}=4$ solutions for which M$_8$ decomposes as $\text{S}^2 \times \text{CY}_2 \times \Sigma_2$. This includes the AdS$_3$ Hopf fiber T-dual of the $\text{CY}_2$ class of \cite{Lozano:2019emq}, studied in \cite{Lozano:2020txg}, and the double analytic continuation of \cite{Lozano:2020txg} studied in \cite{Lozano:2021rmk}. 

In this section we fix the inverse AdS$_2$ radius $m=1$, as we are free to do without loss of generality.

\subsection{\texorpdfstring{${\cal N}=1$ conformal weak G$_2$-holonomy class}{${\cal N}=1$ conformal weak G$_2$-holonomy class}}\label{sec:IIAexample}
In this section we would like to explore the possibility of solutions with internal space decomposing as a weak G$_2$-manifold foliated over an interval. We will thus take the internal metric to decompose as
\beq
ds^2= e^{2C}ds^2(\text{M}_{\text{WG}_2})+ e^{2k} d\rho^2,
\eeq
where $(e^C,e^k)$, the dilaton and the AdS$_2$ warp factor $e^A$ are functions of $\rho$ alone, which the metric of $\text{M}_{\text{WG}_2}$, the weak G$_2$-holonomy manifold, is independent of. A manifold with weak G$_2$-holonomy is 
characterised by the G$_2$-structure 3-form, $\Phi_{\text{WG}_2}$ with a single non-vanishing torsion class, namely
\beq
d\Phi_{\text{WG}_2}= 4 \star_{\text{WG}_2}\Phi_{\text{WG}_2}.
\eeq
Well known examples include G$_2$ cones over nearly-K\"{a}hler bases, the known (closed form) examples of such 6-manifolds are $(\text{S}^6,\,\text{S}^3\times \text{S}^3,\,\mathbb{CP}^3,\,\mathbb{F}^3)$, and one can arrange for compact M$_{\text{WG}_2}$ by fixing
\begin{align}
\Phi_{\text{WG}_2}&=\sin^2\alpha d\alpha\wedge J_{\text{NK}}+\sin^3\alpha \text{Re}(e^{-i\alpha}\Omega_{\text{NK}}),\nn\\
\star_{\text{WG}_2}\Phi_{\text{WG}_2}&= -\frac{1}{2}\sin^4 \alpha J^2_{\text{NK}}+ \sin^3\alpha d\alpha\wedge \text{Im}(e^{-i\alpha}\Omega_{\text{NK}}),
\end{align}
where 
\beq
dJ_{\text{NK}}=3 \text{Re}\Omega_{\text{NK}},~~~~d\text{Im}\Omega_{\text{NK}}=-2J_{\text{NK}}^2,~~~~ds^2(\text{M}_{\text{WG}_2})=d\alpha^2+\sin^2\alpha ds^2(\text{M}_{\text{NK}}).
\eeq
This makes  M$_{\text{WG}_2}$ a folation of a nearly-K\"{a}hler manifold over the interval spanned by $\alpha$, generically tending to a G$_2$ cone singularity\footnote{Such singularities is believed to be allowed in string theory.} at $\alpha=0\sim 2\pi$. The exception is when the base is taken to be S$^6$, in which case M$_{\text{WG}_2}=\text{S}^7$ which is smooth.

Moving forward we shall decompose the bispinors of section \ref{sec:parameterisation} such that\footnote{The presence of a weak G$_2$-manifold does not mean that we must align its associated G$_2$-structure forms along those of section \ref{sec:parameterisation}; this is an assumption. See section 6.2 of \cite{Legramandi:2020txf} for an example that does not conform to this assumption. }
\begin{align}
(a,b)&=(1,0),~~~~\beta=\beta(\rho),~~~~V= e^k d\rho\nn\\
\psi^{(7)}_+&=1+ e^{4C}\star_{\text{WG}_2}\Phi_{\text{WG}_2},~~~~\psi^{(7)}_-=- e^{3C}\Phi_{\text{WG}_2}-e^{7C}\text{vol}(\text{M}_{\text{WG}_2}).
\end{align}
We shall also assume that the RR and NSNS flux preserve the structure of M$_{\text{WG}_2}$, or in other words that the components of $H_3$ and $f_{\pm}$ are non-trivial only along
\beq
d\rho,~~~~\Phi_{\text{WG}_2},~~~~\star_{\text{WG}_2}\Phi_{\text{WG}_2},~~~~\text{vol}(\text{M}_{\text{WG}_2})
\eeq
and wedge products thereof. In particular this forces $H_3=0$ if we want it to obey its Bianchi identity.

Under the above assumptions it is quick to realise that no solutions in Type IIB are possible because the right-hand side of \eqref{BPS2} contains a $\star_{\text{WG}_2}\Phi_{\text{WG}_2}$ that is orthogonal to everything on the left-hand side and cannot be set to zero.  In Type IIA things are better. We decompose the  magnetic  fluxes as
\beq
f_+=  F_0+ e^k p \Phi_{\text{WG}_2}\wedge d\rho+g \star_{\text{WG}_2}\Phi_{\text{WG}_2}+ e^k q \text{vol}(\text{M}_{\text{WG}_2})\wedge d\rho,~~~~H_3=0,\label{eq:RRfluxesweakG2}
\eeq
where $(p,q,g)$ are functions of $\rho$.  The Bianchi identities of the fluxes in regular regions of the space then reduce to 
\beq
dF_0=0,~~~~\partial_{\rho}g+ 4 e^kp=0\label{eq:bianchiweakg2}.
\eeq
Plugging the above Ansatz into \eqref{BPS1}-\eqref{BPS8dpairing} we find it fixes the functions in the flux as
\begin{align}
e^{A+k}\sin\beta F_0&= \partial_{\rho}(e^{A-\Phi}),~~~~e^{A+k}\sin\beta g=\partial_{\rho}(e^{A+4C-\Phi})-4 e^{A+3C+k-\Phi}\cos\beta,\nn\\
e^{A+C+k}\sin\beta p&=-\cos\beta\partial_{\rho}(e^{A+4C-\Phi})+ e^{3C+k-\Phi}(4 e^A+  e^C \sin\beta),\nn\\
e^{A+k-7C}\sin\beta q&=-\cos\beta \partial_{\rho}(e^{A-\Phi})+  e^{k-\Phi}\sin\beta
\end{align}
and furnish us with the following ODEs to solve
\begin{align}
&\partial_{\rho}(e^{3A+7C-2\Phi}\sin\beta)+2 \cos\beta e^{2A+7C+k-2\Phi}\cos\beta=0,\nn\\
&\partial_{\rho}\left(\frac{e^{\frac{7}{2}C-\Phi}\cot\beta}{\sqrt{e^A\sin\beta}}\right)-\frac{e^{\frac{5}{2}C+k-\Phi}}{(e^A\sin\beta)^{\frac{3}{2}}}(3 e^C+14 e^A\sin\beta)=0.\label{eq:waekg2bps}
\end{align}
To solve these we find it useful to introduce two functions of $\rho$, $(G,h)$, defined as
\beq
e^{3A+7C-2\Phi}\sin\beta=-\frac{L^4 h^2}{32 c_0},~~~~ e^{A+7C-2\Phi}\cot\beta=-\frac{L^2 G h}{\sqrt{8} c_0},
\eeq
where $(c_0,L)$ are constants. We then use a coordinate transformation to fix
\beq
e^{A+k}\sin\beta=-\frac{L^2}{8\sqrt{2}},
\eeq
and introduce a function $v=v({\rho})$ such that
\beq
 e^{C}(v-1)=4\sin\beta e^A .
\eeq
This reduces \eqref{eq:waekg2bps} to the rather simple conditions
\beq
G= \frac{1}{c_0 }\partial_{\rho}h,~~~~\cos^2\beta= \frac{c_0  G^2(1-7 v)}{2 h \partial_{\rho}G},
\eeq
which we can use to define $(\cos\beta,G)$ (we take the positive root for $\cos\beta$) thereby solving the necessary conditions for supersymmetry. One can additionally show that all of \eqref{eq:EOMconds} are implied by \eqref{eq:bianchiweakg2}, so finding a solutions boils down to solving these constraints.

In summary, we have found a class of solutions with the following NSNS sector
\begin{align}
\frac{ds^2}{L^2}&=\sqrt{\frac{h}{h''}}\bigg[\frac{h h''\sqrt{1-7v}}{8\Delta}ds^2(\text{AdS}_2)+ \bigg(\frac{h''}{8h\sqrt{1-7v}}d{\rho}^2+\frac{\sqrt{1-7v}}{(v-1)^2}ds^2(\text{M}_{\text{WG}_2})\bigg)\bigg],\nn\\
H&=\frac{L^2}{8\sqrt{2}}d\left(\frac{hh'(1-7v)}{\Delta}-\rho\right)\wedge\text{vol}(\text{AdS}_2),~~~~e^{-\Phi}= \frac{\sqrt{\Delta}(1-v)^{\frac{7}{2}}}{c_0 L^3 (1-7v)^{\frac{5}{4}}}\left(\frac{h''}{h}\right)^{\frac{3}{4}},\nn\\
\Delta&=2h h''-(1-7v)(h')^2,
\end{align} 
where $h'=\partial_{\rho}h$ and the RR sector is defined as in \eqref{eq:RRfluxesweakG2}, where the functions that appear now take the form
\begin{align}
F_0&=\frac{4}{c_0L^4}\partial_{\rho}\left(\frac{(1-v)^{\frac{7}{2}}h''}{(1-7 v)}\right),~~~~g= \frac{4\sqrt{v}}{c_0}\partial_{\rho}\left(\frac{h \sqrt{v}}{\sqrt{1-v}}\right),\\
e^kp&=-\frac{2v^{\frac{3}{4}}}{c_0(1-v)(1-7v)}\partial_{\rho}\left((1-v)^{\frac{3}{2}}v^{\frac{1}{4}}h'\right),~~~~
e^kq=\frac{L^4}{c_0}\left(\partial_{\rho}\left(\frac{(1-7 v)h h'}{(1-v)^{\frac{7}{2}}h''}\right)-\frac{h (3-7 v)}{(1-v)^{\frac{7}{2}}}\right)\nn.
\end{align}
The Bianchi identities \eqref{eq:bianchiweakg2},  away from sources reduce to
\begin{align}
&\partial^2_{\rho}\left(\frac{(1-v)^{\frac{7}{2}}h''}{(1-7 v)}\right)=0,\nn\\
&\partial_{\rho}\left(\sqrt{v}\partial_{\rho}\left(\frac{h \sqrt{v}}{\sqrt{1-v}}\right)\right)-\frac{2v^{\frac{3}{4}}}{(1-v)(1-7v)}\partial_{\rho}\left((1-v)^{\frac{3}{2}}v^{\frac{1}{4}}h'\right)=0,\label{eq:wg2odes}
\end{align}
and one has an (at least) ${\cal N}=1$ supersymmetric solution whenever these are solved.

The conditions \eqref{eq:wg2odes} appear rather difficult to solve in full generality. One could proceed semi-analytically, i.e.\  solve the above with a series expansion and then attempt to numerically interpolate; we will not attempt this here. Instead we note that when one fixes $v=v_0$, a constant, \eqref{eq:wg2odes} reduces to
\beq
h''''=0,~~~~ v_0(1+5 v_0)=0,
\eeq
with both $v_0=0$ and $1+5 v_0=0$ being valid independent solutions. In each case we have  $h'''\propto F_0$, so $h$ is locally a degree three polynomial, though $F_0$ need not be fixed globally.  This fact can potentially be used to glue local patches with different values of $F_0$ together, leading  to infinite classes of solution with D8-branes at the loci where $F_0$ is discontinuous. Such behaviour has been observed and exploited before in the context of AdS$_{7,3,2}$ solutions of massive Type IIA \cite{Apruzzi:2013yva,Lozano:2022ouq,Macpherson:2023cbl,Dibitetto:2018gbk}, the last being the most relevant to the case at hand.

When $v_0=0$ the only non-trivial magnetic RR fluxes are  $f_0$ and $f_8$, and by comparing to \cite{Dibitetto:2018gbk} it becomes clear we have a generalisation of the class of ${\cal N}=8$ AdS$_2\times \text{S}^7\times \mathcal{I}$ solutions found there, where S$^7$ can now be any weak G$_2$-manifold and generically only ${\cal N}=1$ is preserved. That such solutions exist is not surprising, as S$^7$ indeed supports a weak G$_2$-holonomy. 

When $1+5 v_0=0$ the metric gets deformed with respect to the $v_0=0$ case, now all of $(f_0,f_4,f_8)$ are non-trivial, but solutions are still governed by the same ODE, $h''''=0$. It would be interesting to explore what solutions lie within this class, but that is beyond the scope of our aims here. We note that the ${\cal N}=8$ AdS$_2$ solutions of \cite{Dibitetto:2018gbk} can be mapped to the AdS$_7$ solutions of \cite{Apruzzi:2013yva} via double analytic continuation. It would also be interesting to explore the class of deformed (with regard to the fluxes only) AdS$_7$ solutions that the $1+5 v_0=0$ case should map to, and what implications they could have in the context of the AdS$_7$/CFT$_6$ correspondence.

\subsection{\texorpdfstring{Small ${\cal N}=4$ $\text{AdS}_2\times\text{S}^2\times\text{CY}_2\times\Sigma_2$ class}{Small ${\cal N}=4$ AdS(2) x S**2 x CY(2) x Sigma(2) class}}\label{sec:IIBexample}
Two examples of  classes of solutions with AdS$_2\times\text{S}^2\times\text{CY}_2\times\text{S}^1$  foliated over an interval and  preserving small ${\cal N}=4$ were derived in \cite{Lozano:2020txg}. The first via AdS$_3$ Hopf fiber T-dual of an AdS$_3$ class derived in \cite{Lozano:2019emq}, the second \cite{Lozano:2021rmk} with a double analytic continuation of the first. In this section we will construct a much broader class of solutions containing both these examples as limits.

We seek a solution for which the internal manifold and magnetic NSNS 3-form  decompose as 
\beq
ds^2= e^{2C}ds^2(\text{S}^2)+ ds^2(\text{M}_4)+ V^2+ U^2,~~~~H_3= e^{2C}\tilde{H}_1\wedge \text{vol}(\text{S}^2)+ \tilde{H}_3,
\eeq
where $(e^{2C},\tilde{H}_1,\tilde{H}_3)$ have support on M$_4$, and where the magnetic component of the RR fluxes decompose in a similar SU$(2)$ invariant fashion. In addition to the SU$(2)$ invariant $\text{vol}(\text{S}^2)$, S$^2$ supports the following triplets in terms of unit radius embedding coordinates $\mu_{\mathfrak{a}}$, $\mathfrak{a}=1,2,3$,
\beq
\mu_\mathfrak{a},~~~~d\mu_\mathfrak{a}~~~~\epsilon_{\mathfrak{a}\mathfrak{b}\mathfrak{c}}\mu_\mathfrak{b}d\mu_\mathfrak{c},~~~~\mu_\mathfrak{a} \text{vol}(\text{S}^2).
\eeq
In terms of these we make the following Ansatz\footnote{This Ansatz is by no means general for $\text{AdS}_2\times \text{S}^2$ solutions, it is merely sufficient to construct the class we seek.} for the SU$(3)$-structure forms and functions appearing in \eqref{eq:su3structure}
\beq
J=  e^{2C}\text{vol}(\text{S}^2)- \mu_\mathfrak{a} j_\mathfrak{a},~~~~\Omega= e^{C}\left(d\mu_\mathfrak{a}\wedge j_\mathfrak{a}+ i \epsilon_{\mathfrak{a}\mathfrak{b}\mathfrak{c}}\mu_\mathfrak{b}d\mu_\mathfrak{c}\wedge j_\mathfrak{c}\right),~~~~a+i b= e^{-i\alpha},\label{eq:IIBexampleforms}
\eeq
where we will assume that $(U,V)$ have no support on the $2$-sphere. Here $j_\mathfrak{a}$ are a set of SU$(2)$-structure 2-forms on M$_4$ obeying
\beq
j_\mathfrak{a}\wedge j_\mathfrak{b}= 2 \delta_{\mathfrak{a}\mathfrak{b}}\text{vol}(\text{M}_4).
\eeq
This choice of SU$(3)$-structure forms ensures that we have ${\cal N}=4$ supersymmetry provided that the fluxes are SU$(2)$ singlets. This is because we are free to rotate $\mu_\mathfrak{a}$ by a constant element of SO$(3)$ and change nothing about the physical fields. It is clear from  \eqref{eq:IIBexampleforms} that only the charged parts of \eqref{eq:IIBexampleforms} transform under this action so if we demand that the fluxes are SU$(2)$ singlets one can generate further three independent SU$(3)$-structures in this fashion yielding the claimed ${\cal N}=4$.  The necessary conditions for supersymmetry \eqref{BPS2}-\eqref{BPS3} then decompose into singlet and triplet parts under SU$(2)$. We want the RR flux to preserve this symmetry so only the former may contribute to this. After a long computations one  can show that  fixing the triplet parts of \eqref{BPS2}-\eqref{BPS3} to zero amounts to imposing
\begin{subequations}
\begin{align}
&e^C+ \sin\beta\sin\alpha   e^A=0,\label{eq:bpsIIB1}\\
&e^{2C}\tilde{H}_1+ e^{A}\sin\beta U+ d\left(e^{2A}\cos\alpha\sin\beta\right)=0,\label{eq:bpsIIB2}\\
&d(e^{A}\sin\beta V)=0,\label{eq:bpsIIB3}\\
&d(e^{A}\sin\beta U)\wedge j_\mathfrak{a}=0,\label{eq:bpsIIB4}\\
&d(e^{2A-\Phi} \sin\alpha\sin\beta  j_\mathfrak{a})-e^{2A-\Phi}(\cos\alpha U-\cos\beta \sin\alpha V)\wedge j_\mathfrak{a}=0,\label{eq:bpsIIB5}\\
&\tilde{H}_3\wedge j_\mathfrak{a}=\tilde{H}_3\wedge U\wedge V=0\label{eq:bpsIIB6}.
\end{align}
\end{subequations}
The first two of these just define a warp factor and part of the NSNS flux, it is the rest that we must solve for. First, we note that $d(e^{2C}\tilde{H}_1)=0$ is a necessary condition for the Bianchi identity of the NSNS flux away from the loci of sources, as such in regular regions of a solution we have that  $d(e^{A}\sin\beta U)=0$, which implies \eqref{eq:bpsIIB4}. We solve this condition and \eqref{eq:bpsIIB3} in terms of two local coordinates $(x_1,x_2)$ as
\beq
U+i V=-\frac{e^{-A}}{\sin\beta}(dx_1+i dx_2).
\eeq
Taking the exterior derivative of \eqref{eq:bpsIIB5}, it then follows that
\beq
dx_2\wedge d(e^{-2A}\cot\alpha\csc^2\beta)= dx_1\wedge d(e^{-2A}\cos\beta \csc^2\beta),
\eeq
which ensures that the combinations in parentheses are functions of $(x_1,x_2)$ alone, and obey
\beq
\partial_{x_1}(e^{-2A}\cot\alpha\csc^2\beta)+\partial_{x_2}(e^{-2A}\cos\beta \csc^2\beta)=0.
\eeq
This is an integrability condition which we are free to solve by introducing a local function $u=u(x_1,x_2)$ such that 
\beq
e^{-2A}\cot\alpha\csc^2\beta=-\partial_{x_2}\log u,~~~~e^{-2A}\cos\beta \csc^2\beta=\partial_{x_1}\log u.
\eeq
It is then possible to recast  \eqref{eq:bpsIIB4} in the form
\beq
d\left(\frac{e^{2A-\Phi}\sin\alpha\sin\beta}{u}j_{\mathfrak{a}}\right)=0,
\eeq
which informs us that M$_4$ is conformally a Calabi-Yau $2$-fold, specifically we have
\beq
ds^2(\text{M}_4)=\frac{e^{-2A+\Phi} u }{\sin\alpha \sin\beta}ds^2(\text{CY}_2).
\eeq
Finally, the condition \eqref{eq:bpsIIB6} informs us that $\tilde{H}_3$ must decompose in terms of two primitive $(1,1)$-forms on $\text{CY}_2$ as\footnote{In this instance a primitive $(1,1)$-form $X^{(1,1)}$ is defined such that it has legs only in the CY$_2$ directions and satisfies $X^{(1,1)}\wedge j_\mathfrak{a}=0$. In a canonical frame on CY$_2$ where $(j_3=\e^{12}+\e^{34},~j_1+i j_2=(\e^1+i \e^2)\wedge (\e^3+i \e^4))$ this means that $X^{(1,1)}$ decomposes in a basis of three independent forms $(\e^{12}-\e^{34},\,\e^{13}+\e^{24},\,\e^{14}-\e^{23})$, as such each of $X^{(1,2)}_{1,2}$ can depend on three independent functions of $(x_1,x_2,\text{CY}_2)$.}
\beq
\tilde{H}_3= dx_1\wedge X_1^{(1,1)}+dx_2\wedge X_2^{(1,1)},
\eeq
with $(X_1^{(1,1)},\,X_2^{(1,1)})$ anti-self-dual on CY$_2$ by definition. 
Having solved the triplet contributions of \eqref{BPS2}-\eqref{BPS3} we can now extract the RR flux from \eqref{BPS3} as in \eqref{eq:flux def}, and substitute the result into \eqref{BPS2} and \eqref{BPS3}. The first of these results in a single additional condition 
\beq
d(e^{-\Phi}\sin^2\alpha\csc\beta)\wedge dx_1\wedge dx_2=0,
\eeq
which informs us that $e^{-\Phi}\sin^2\alpha\csc\beta$ is independent of the coordinates on CY$_2$. We solve this condition in terms of another local function $h_7=h_7(x_1,x_2)$ and a constant $c_0$ as
\beq
\frac{e^{-\Phi}\sin\alpha}{\sin\beta}=c_0 h_7.
\eeq
Due to our judicious redefinitions of the physical fields, \eqref{BPS3} then reduces to simply
\beq
\nabla_2^2 u=0,~~~~\nabla_2^2:= \partial_{x_1}^2+\partial_{x_2}^2.
\eeq
As \eqref{BPS1} can be taken to define the electric part of the NSNS flux, the necessary conditions for supersymmetry are now solved. However, we find it useful to introduce one final local function $h_3=h_3(x_1,x_2,\text{CY}_2)$ as
\beq
\frac{u^2}{e^{4A}h_7\sin^4\beta}= h_3,
\eeq
before we summarise our results and present the explicit form of the RR fluxes.

In summary, we find a class of solutions with NSNS sector of the form
\begin{align}
ds^2&=  \frac{u}{\sqrt{h_3 h_7}}\left(\frac{1}{\Delta_2}ds^2(\text{AdS}_2)+\frac{1}{\Delta_1}ds^2(\text{S}^2)\right)+\sqrt{\frac{h_3}{h_7}}ds^2(\text{CY}_2)+\frac{\sqrt{h_3h_7}}{u}(dx_1^2+dx_2^2)\bigg],\nn\\
e^{-\Phi}&=c_0 \sqrt{\Delta_1\Delta_2} h_7,~~~~ H= dB_0\wedge \text{vol}(\text{AdS}_2)+d\tilde{B}_0\wedge\text{vol}(\text{S}^2)+ dx_1\wedge X^{(1,1)}_1+ dx_2\wedge X^{(1,1)}_2,\nn\\
\Delta_1&=1+ \frac{(\partial_{x_1}u)^2}{h_3 h_7},~~~~\Delta_2=1- \frac{(\partial_{x_2}u)^2}{h_3 h_7},~~~~B_0=-x_2-\frac{u \partial_{x_2}u}{ h_3 h_7\Delta_2},~~~~\tilde{B}_0=x_1-\frac{u \partial_{x_1}u}{ h_3 h_7\Delta_1},
\end{align}
for functions $(u,h_7)$ with support on $(x_1,x_2)$ and $h_7$ with support on $(x_1,x_2,\text{CY}_2)$. Solutions in this class support the following non-trivial magnetic RR fluxes\footnote{The electric components are defined in terms of these as in \eqref{eq:AdS2vacua}.
}
\begin{align}
f_1&=c_0\left(\star_2 d_2h_7+d\left(\frac{\partial_{x_1}u\partial_{x_2}u}{h_3}\right)\right),\nn\\
f_3&= \tilde{B}_0f_1\wedge \text{vol}(\text{S}^2)-c_0\bigg[\bigg(x_1\star_2 d_2h_7+ h_7dx_2-d\left(\frac{\partial_{x_2}u(u-x_1\partial_{x_1}u)}{h_3}\right)\bigg)\wedge\text{vol}(\text{S}^2)\nn\\
&+h_7\left(dx_1\wedge X^{(1,1)}_2-dx_2\wedge X^{(1,1)}_1\right)+\frac{\partial_{x_1}u\partial_{x_2}u}{h_3}\left( dx_1\wedge X^{(1,1)}_1+ dx_2\wedge X^{(1,1)}_2\right)\bigg],\nn\\
f_5&= \tilde{B}_0f_3\wedge \text{vol}(\text{S}^2)-c_0\bigg[\frac{h_7}{u}\star_4 d_4 h_3\wedge \text{vol}_2+ \star_2 d_2 h_3\wedge\text{vol}_4+ d\left(\frac{\partial_{x_1}u\partial_{x_2}u}{h_7}\right)\wedge \text{vol}_4\nn\\
&+\bigg(-x_1 h_7\left(dx_1\wedge X_2^{(1,1)}-dx_2\wedge X_1^{(1,1)}\right)+\frac{\partial_{x_2}u(u-x_1 \partial_{x_1}u)}{h_3}\left(dx_1\wedge X_1^{(1,1)}+dx_2\wedge X_2^{(1,1)}\right)\bigg)\wedge \text{vol}(\text{S}^2)\bigg],\nn\\
f_7&= \tilde{B}_0f_5\wedge \text{vol}(\text{S}^2)+c_0\bigg[-d\left(\frac{\partial_{x_2}u(u-x_1\partial_{x_1}u)}{h_7}\right)\wedge \text{vol}_4\nn\\
&+x_1\frac{h_7}{u}\star_4 d_4 h_3\wedge \text{vol}_2+(x_1 \star_2d_2 h_3+h_3 dx_2)\wedge \text{vol}_4\bigg]\wedge \text{vol}(\text{S}^2),
\end{align}
where we have decomposed
\beq
d=d_2+ d_{4},~~~~ d_2= dx_i\wedge  \partial_{x_i},
\eeq
for $d_4$ the exterior derivative on CY$_2$ and where $\star_{2,4}$ and $\text{vol}_{2,4}$ are the Hodge duals and volume forms on the unwarped $(x_1,x_2)$ and CY$_2$ sub-manifolds respectively. 

Supersymmetry is solved by simply imposing that globally
\beq
\nabla_2^2u=0,
\eeq
i.e.\ $u$ must be a harmonic function on $(x_1,x_2)$, and that $(X^{(1,1)}_1,\, X^{(1,1)}_2)$ are primitive $(1,1)$-forms with ``legs'' on CY$_2$ only. Thus, finding solutions becomes a two step process: first one chooses a harmonic function $u$ and specific CY$_2$ manifold and then one needs to solve \eqref{eq:EOMconds} for this choice. The Bianchi identities of the magnetic fluxes impose that in regular regions of the internal space
\begin{align}
&d_4 X^{(1,1)}_1=d_4 X^{(1,1)}_2=0,~~~~\partial_{x_2}X^{(1,1)}_1=\partial_{x_1}X^{(1,1)}_2,~~~~\partial_{x_1}(h_7^2X^{(1,1)}_1)=-\partial_{x_2}(h_7^2X^{(1,1)}_2),\nn\\
& \nabla_2^2 h_7=0,~~~~  \frac{h_7}{u}\nabla_4^2h_3+\nabla_2^2 h_3+ h_7\left((X^{(1,1)}_1)^2+(X^{(1,1)}_2)^2\right)=0,\label{eq:IIBexamplebianchi}
\end{align}
where $\nabla_{2,4}^2$ are the Laplacians on the unwarped $(x_1,x_2)$ and CY$_2$ directions.  When these conditions hold, all of \eqref{eq:EOMconds} are implied so one has a solution.

To better understand this class it is instructive to fix  $X^{(1,1)}_1=X^{(1,1)}_2=0$ and $u=1$ (only $u=$ constant is required but one can then fix $u=1$ without loss of generality by rescaling coordinates and the other functions). First off, in this limit \eqref{eq:IIBexamplebianchi} can be solved by fixing both $(h_3,h_7)$ to be constant, we then recover the double T-dual of the D1-D5 near-horizon performed on the Hopf fiber of both its AdS$_3$ and S$^3$ factors. If we allow for non-constant $(h_3,h_7)$  then  \eqref{eq:IIBexamplebianchi} reduce to the Bianchi identities on would expect for  D3-branes localised within the worldvolume of D7-branes whose relative codimensions are a CY$_2$ manifold. The metric and dilaton also take the expected form for such branes if they are both extended in AdS$_2\times$S$^2$, as such, at least in this limit  $(h_3,h_7)$ play the role of (D3,D7)-brane warp factors which is the reason for their numerical subscripts.  If we now turn on  $X^{(1,1)}_{1,2}$ the metric remains unchanged but the D3-brane PDE inherits an additional term from the fluxes. Allowing $u$ to be a more general harmonic function deforms this system and the interpretation of $(h_3,h_7)$ in terms of warp factors of branes is not at all clear, indeed this likely depends on the specific choice of $u$.

If one imposes that $\partial_{x_1}$ spans a U$(1)$ isometry and additionally fixes $X^{(1,1)}_1=0$ we recover the class of solutions in \cite{Lozano:2021rmk} obtained by double analytic continuation of the class in \cite{Lozano:2020txg}.
If instead we make $\partial_{x_2}$ a U$(1)$ isometry  and $X^{(1,1)}_2=0$ one recovers the result of T-dualising the $\text{AdS}_3\times \text{S}^2\times \text{CY}_2\times \mathcal{I}$ class of solutions found in \cite{Lozano:2019emq} on the Hopf fiber of AdS$_3$, as studied in \cite{Lozano:2020txg}. One can of course perform this T-duality in the presence of a non-trivial $X^{(1,1)}_2$, in which case AdS$_3$ becomes fibred over the internal CY$_2$ generalising the class in \cite{Lozano:2019emq}. If we additionally impose that $\partial_{x_1}$ is an isometry we further generalise the generalised D1-D5 near-horizon geometries studied in \cite{Lima:2022hji} such that both the AdS$_3$ and S$^3$ factors are fibered over $\text{CY}_2$.

Specific solutions to the classes in \cite{Lozano:2020txg} and \cite{Lozano:2021rmk} were considered in the limit that the symmetries of the CY$_2$ were respected, i.e.\ with fluxes only depending on the CY$_2$ through $\text{vol}_4$ and where the warp factors only depend on $(x_1,x_2)$. In this limit, the classes of \cite{Lozano:2020txg,Lozano:2021rmk} can be embedded into a classification of $\text{AdS}_2\times \text{S}^2\times \text{CY}_2\times \Sigma_2$ in \cite{Chiodaroli:2009yw,Chiodaroli:2009xh}, which likewise restricts dependence on CY$_2$. It would be interesting to see how the $X^{(1,1)}_1=X^{(1,1)}_2=0$ and $h_3=h_3(x_1,x_2)$ limit of the class we present here fits within this existing classification, and whether it is actually equivalent. We stress that as our class has non-trivial $(X^{(1,1)}_1,\,X^{(1,1)}_2)$ and $h_3=h_3(x_1,x_2,\text{CY}_2)$ it lies beyond the results of \cite{Chiodaroli:2009yw,Chiodaroli:2009xh}.

In the limit $u=1$ we note that AdS$_2$ and S$^2$ share the same warping, so this class could potentially be fruitful in the study of four-dimensional black hole near-horizon geometries. We leave this interesting possibility to be studied elsewhere.

\section*{Acknowledgements}
We thank Yolanda Lozano, Carlos Nunez and  Alessandro Tomasiello for comments on a draft of this work. We also thank Alessandro Tomasiello for several illuminating discussions during this project. The work of NM is supported by the Ram\'on y Cajal fellowship RYC2021-033794-I, and by grants from the Spanish government MCIU-22-PID2021-123021NB-I00 and principality of Asturias SV-PA-21-AYUD/2021/52177. The work of AP was funded, in whole, by ANR (Agence Nationale de la Recherche) of France under contract number ANR-22-ERCS-0009-01. AL acknowledges support from Provincia Autonoma di Trento and by Q@TN, the joint lab between University of
Trento. This project has received funding from European Research Council (ERC) under the European Union’s Horizon 2020 research and innovation programme (grant agreement No 804305).

\appendix

\section{Type II supergravity and conventions}\label{sec:convensions}
We work with the ``democratic'' conventions of \cite{Tomasiello:2011eb} for Type II supergravities. The bosonic fields split into two sectors: the NSNS sector containing the metric, dilaton and NSNS flux, respectively $(g,\Phi,H)$, and the RR sector containing the RR polyform $F$. The RR flux is subject to the self-duality constraint
\beq
F =  \star \lambda(F),
\eeq
where for a $k$-form $C_k$ --- to which we will make frequent reference in this section ---  $\lambda(C_k) := (-)^{\lfloor\frac{k}{2}\rfloor} C_k$, and we define Hodge dualisation in all dimensions and signatures as
\beq
\star \e^{\, \underline{M}_1\dots\underline{M}_k}:=  \frac{1}{(d-k)!}\epsilon_{\underline{M}_{k+1}\dots\underline{M}_{d-k}}{}^{\underline{M}_1\dots\underline{M}_k} \, \, \e^{\, \underline{M}_{k+1}\dots\underline{M}_{d-k}},
\eeq
for $\e^{\underline{M}}$ a vielbein (underlined indices are flat spacetime indices).

A solution to Type II supergravity must obey the following equations of motion and Bianchi identities away from the loci of sources
\begin{align}
&d_H F=0,~~~~ dH=0,~~~~ d(e^{-2\Phi}\star H)=\frac{1}{2}(F,F)_8,
\nn \\
& 2R- H^2-8 e^{\Phi}\nabla^2 e^{-\Phi}=0,~~~~  R_{MN}+2 \nabla_{M}\nabla_{N}\Phi-\frac{1}{2} H^2_{MN}-\frac{e^{2\Phi}}{4} (F)^2_{MN}=0,
\end{align}
where $(F,F)_8$ is the 8-form part of $F\wedge \lambda(F)$, and 
\begin{align}
(C_k)_{M}&:= \iota_{dx^M} C_k,~~~~ (C_k)^2:= \sum_{k} \frac{1}{k!} (C_k)_{M_1\dots M_k}(C_k)^{M_1\dots M_k},~~~~\nn\\
C^2_{MN}&:=\sum_{k} \frac{1}{(k-1)!} (C_k)_{MM_1\dots M_{k-1}}(C_k)_N{}^{M_1\dots M_{k-1}}.
\end{align}
Such a solution preserves supersymmetry if it supports two Majorana--Weyl spinors $\epsilon_{1,2}$ that satisfy the constraints (here and elsewhere the upper/lower signs are taken in Type IIA/IIB supergravity)
\begin{subequations}
\begin{align}
& \left(\nabla_M-\frac{1}{4}H_M\right) \epsilon_1 + \frac{e^\Phi}{16}F \Gamma_M\epsilon_2,~~~~ \left(\nabla_M+\frac{1}{4}H_M\right) \epsilon_2 \pm \frac{e^\Phi}{16} \lambda(F) \Gamma_M \epsilon^1=0, \label{eq:10dsusysseqs1}\\
& \left(\nabla-\frac{1}{4}H-d\Phi\right)\epsilon_1=0,~~~~
\left(\nabla+\frac{1}{4}H-d\Phi\right)\epsilon_2=0,\label{eq:10dsusysseqs4}
\end{align}
\end{subequations}
where here and throughout this work when a form acts on a spinor it does so through the Clifford map, i.e.\   
\beq\label{eq:actiondef}
C_k\epsilon= \slashed{C_k}\epsilon :=  (C_k)^{M_1\dots M_k}\Gamma_{M_1\dots M_k}\epsilon.
\eeq
More generally the Clifford map states that the following are equivalent
\beq
C_k =  (C_k)_{\underline{M}_1\dots \underline{M}_k}\e^{\, \underline{M}_1\dots \underline{M}_k} \equiv  \slashed{C_k} =  (C_k)_{\underline{M}_1\dots \underline{M}_k}\Gamma^{\underline{M}_1\dots \underline{M}_k},
\eeq
where $\Gamma^{\underline{M}}$ is a basis of flat spacetime gamma matrices. This implies the following equivalence for spinor bilinears in $d$ even dimensions 
\beq
\epsilon_1\otimes \epsilon_2^{\dag}= \frac{1}{2^{\lfloor\frac{d}{2}\rfloor}}\sum_{k=1}^d\frac{1}{k!}\epsilon_2^{\dag}\Gamma_{\underline{M}_k\dots\underline{M}_1}\epsilon_1\Gamma^{\underline{M}_1\dots\underline{M}_k}\equiv\frac{1}{2^{\lfloor\frac{d}{2}\rfloor}}\sum_{k=1}^d\frac{1}{k!}\epsilon_2^{\dag}\Gamma_{\underline{M}_k\dots\underline{M}_1}\epsilon_1\e^{\, \underline{M}_1\dots\underline{M}_k},
\eeq 
in odd dimensions the right-hand side of the above equivalence contains the left-hand side twice so one needs to add a $\frac{1}{2}$ to account for this. 
The action of gamma matrices on forms can then be viewed through the following map
\beq
\Gamma^{\underline{M}}  C_k= (\e^{\underline{M}}\wedge +\iota_{\e^{\underline{M}}})C_k,~~~~ (-)^k C_k\Gamma^{\underline{M}} = (\e^{\underline{M}}\wedge -\iota_{\e^{\underline{M}}})C_k,
\eeq
where strictly speaking these expressions hold inside a slash as in \eqref{eq:actiondef}, but we shall suppress this.

The basis of gamma matrices appearing here is such that, for an intertwiner $B^{(10)}$, defining Majorana conjugation as  $\epsilon^c := B^{(10)}\epsilon^*$  we have
\beq
(B^{(10)})^{-1}\Gamma_M B^{(10)}= \Gamma^*_{M},~~~~ B^{(10)} (B^{(10)})^*=\mathbb{I}.
\eeq
We define the chirality matrix in ten dimensions to be 
\beq
\hat\Gamma:= \Gamma^{\underline{0}}\dots\Gamma^{\underline{9}}
\eeq
and the spinors obey the following under it
\beq
\hat\Gamma\epsilon_1= \epsilon_1,~~~~ \hat\Gamma\epsilon_2=\mp \epsilon_2.
\eeq
Our conventions thus far imply another useful relationship
\beq
\hat \Gamma C_k=  \star \lambda(C_k),
\eeq
where again we leave the slash implicit.

\section{\texorpdfstring{Killing spinors and Killing spinor bilinears on AdS$_2$}{Killing spinors and Killing spinor bilinears on AdS(2)}}\label{sec:appenixAdS2bilinears}
On  AdS$_2$ of inverse radius $m$, there are Killing spinors of $\pm$ chirality $\zeta_{\pm}$ that obey the Killing spinor equation
\beq\label{eq:ADS2KSE}
\nabla_{\mu}\zeta_{\pm}=\frac{m}{2}\gamma^{(2)}_{\mu}\zeta_{\mp}.
\eeq
We can take $\gamma^{(2)}_{\mu}$ to be real, so that $\zeta_{\pm}$ are also real. One can then show that the spinors give rise to the following bilinears under the Clifford map
\beq
\zeta_{\pm}\otimes \overline{\zeta}_{\pm}= \frac{1}{2}(v\pm u),~~~~\zeta_{\pm}\otimes \overline{\zeta}_{\mp}=\frac{1}{2}f (\pm 1- \text{vol}(\text{AdS}_2)),
\eeq
where $\bar{\zeta} := (\gamma^{(2)}_{\underline{0}} \zeta )^{\dag}$, and these objects obey the following differential relations
\beq
df=-m u,~~~~ dv=-2mf \text{vol}(\text{AdS}_2),~~~~\nabla_{(\mu}v_{\nu)}=0,~~~~\nabla_{(\mu}u_{\nu)}=-m f g_{\mu\nu}.
\eeq
Thus $v^{\mu}\partial_{x^{\mu}}$ is a Killing vector, while  $u^{\mu}\partial_{x^{\mu}}$ is not, rather it is a conformal Killing vector. The forms also obey the following conditions
\begin{align}
& |v|^2=- f^2,~~~~|u|^2=f^2,~~~~u\cdot v=0,\nn\\
&\iota_u \text{vol}(\text{AdS}_2)=v,~~~~\iota_v \text{vol}(\text{AdS}_2)=u,~~~~ f^2\text{vol}(\text{AdS}_2)=- v\wedge u.
\end{align}
For the pairing constraints some additional conditions will be useful; one can show that
\beq
v\zeta_{\pm}= \pm f  \zeta_{\mp},~~~~u\zeta_{\pm}=-f \zeta_{\mp}.
\eeq
Thus $v^{\pm}= \frac{1}{2} (v\pm u)$ is such that
\beq
v^{\pm}\zeta_{\pm}=0,~~~~v^{\mp}\zeta_{\pm}=\pm  f \zeta_{\mp},~~~~\overline{\zeta}_{\pm}v^{\mp}\zeta_{\pm}=2v^{\mp}\cdot v^{\pm} = -  f^2,~~~~ v^{\pm}\cdot v^{\pm}=0.
\eeq
The identities presented in this section are most easily derived in terms of a specific parameterisation of AdS$_2$, for instance one can take
\beq
ds^2(\text{AdS}_2)= - e^{2m r}dt^2+dr^2.
\eeq
Taking the flat spacetime gamma matrices to be $\gamma^{(2)}_{\underline{\mu}}=(i\sigma_2,\sigma_1)_{\underline{\mu}}$, and with respect to the obvious vielbein suggested by the diagonal metric above, we have a $t$-independent solution to \eqref{eq:ADS2KSE} given by:
\beq
\zeta_+=e^{m r}\left(\begin{array}{c}1\\0\end{array}\right),~~~~\zeta_-=e^{m r}\left(\begin{array}{c}0\\1\end{array}\right) \label{eq:AdSspinorrep}.
\eeq
In terms of these we then have
\beq
v= e^{2m r}dt,~~~~ u= -e^{m r} dr,~~~~f= e^{m r}.
\eeq
Of course there is a second solution to \eqref{eq:ADS2KSE}, dual to conformal supercharges rather than space-time ones under the AdS/CFT. One could choose to align $\zeta_{\pm}$ along a linear combination of these and what appears in  \eqref{eq:AdSspinorrep}. This will lead to a different form of $(f, u,v)$, but they will still obey the identities presented earlier.

\section{Refining the pairing constraints for the time-like Killing vector case}\label{sec:refinment}
Following \cite{Tomasiello:2011eb}, a supersymmetric solution of Type II supergravity must obey the following differential conditions
\beq
d\tilde{K}= \iota_{K} H,~~~~ d_H(e^{-\Phi}\Psi)=- (\iota_{K}+\tilde{K}\wedge )F,~~~~ \nabla_{(M}K_{M)}=0,\label{eq:10dsusy1}
\eeq
where  $(H,F)$ are the NSNS and RR fluxes, $\Phi$ is the dilaton and the remaining objects can be defined in terms of the two Majorana--Weyl Killing spinors $\epsilon_{1,2}$ that Type II solutions can support as
\beq
K:=\frac{1}{2}(K_1+K_2),~~~~ \tilde{K}:=\frac{1}{2}(K_1-K_2),~~~~ K_i:=\frac{1}{32}\overline{\epsilon}_i\Gamma_M\epsilon_i \e^M,~~~~\Psi:= \epsilon_1\otimes \overline{\epsilon}_2,
\eeq
where  $\bar{\epsilon}:= (\Gamma_{\underline{0}} \epsilon )^{\dag}$,  $K_{1,2}$ are  null and obey $K_i\epsilon_i=0$ and where $K^M \partial_M$ is a Killing vector with respect to all bosonic supergravity fields and with respect to which $\epsilon_{1,2}$ are singlets.

The above conditions are not however sufficient to imply supersymmetry in general; for that one must solve the rather more cumbersome pairing constraints, namely
\begin{subequations}
\begin{align}
\left( \e_{1+}\Psi \e_{2+},~\Gamma^{MN}\bigg[(-1)^{\text{deg}(\Psi)} d_H\left(e^{-\Phi}\Psi \e_{+2}\right)+\frac{e^{\Phi}}{2}\star d(e^{-2\Phi}\star \e_{+2})\Psi-F\bigg]\right)=0,\label{eq:10dsusy2}\\
\left( \e_{1+}\Psi \e_{2+},~\bigg[ d_H\left(e^{-\Phi} \e_{+1}\Psi \right)-\frac{e^{\Phi}}{2}\star d(e^{-2\Phi}\star \e_{+1})\Psi-F\bigg]\Gamma^{MN}\right)=0,\label{eq:10dsusy3}
\end{align}
\end{subequations}
where $\e_{1,2+}$ are non-trivial null vectors/1-forms that must  obey
\beq
\e_{1+} \cdot K_{1}=\e_{2+} \cdot K_{2} =\frac{1}{2},
\eeq
but one is otherwise free to choose the precise form of. 
The pairing itself as a geometric object is simply defined as  $(X,Y)_{d}:= X\wedge \lambda (Y)\lvert_{d}$, however it also has equivalent representations in terms of a spinor bilinear and trace, i.e.\ in dimension $d$ we have
\begin{align}
& \frac{1}{2^{[\frac{d}{2}]}}\text{Tr}(\star X Y)= (-1)^{\text{deg}(X)}(X,Y),\nn\\
&(X \Psi Y, C)=-\frac{1}{32}(-1)^{\text{deg}(\Psi)}  \overline{\epsilon}_1 X C Y \epsilon_2\text{vol}_{d},\nn\\
&\overline{\epsilon}_1XC Y\epsilon_2=- (-1)^{\text{deg}(\Psi)}\text{Tr}(Y\lambda (\Psi)X C),\label{eq:pairingidenities}
\end{align}
which we make extensive use of in this and the following appendix. 
The conditions \eqref{eq:10dsusy1} and \eqref{eq:10dsusy2}-\eqref{eq:10dsusy3} are necessary and sufficient for supersymmetry. Let us now simplify these conditions under the assumption that $K^M\partial_M$ is a  time-like Killing vector.

The first thing we need to do is to make a choice for $\e_{1,2+}$.  Given that
\beq
K_1\cdot K_2=2 K^2 \neq 0,
\eeq
in the timelike case we can simply take
\beq
\e_{1+}= \frac{1}{4K^2}K_2,~~~~ \e_{2+}= \frac{1}{4K^2}K_1.
\eeq
It is then simple to show that three of the objects appearing in the pairing constraints can be written as 
\beq
\e_{1+}\Psi \e_{2+}\propto K \Psi K,~~~~\Psi \e_{2+}= \frac{1}{2K^2}\Psi K,~~~~\e_{1+}\Psi = \frac{1}{2K^2}K\Psi 
\eeq
Moving forward, a useful condition was already derived in \cite{Legramandi:2018qkr} namely for time-like,
\beq
d(e^{-2\Phi} \star K_{1,2})=0,
\eeq
importantly these conditions are implied by \eqref{eq:10dsusy1}. From this it follows that
\begin{align}
\frac{e^{\Phi}}{2}\star d(e^{-2\Phi}\star \e_{2+})=-\frac{e^{\Phi}}{2}\star d(e^{-2\Phi}\star \e_{1+}) =\frac{e^{-\Phi}}{8}{\cal L}_{\tilde{K}}K^{-2},
\end{align}

Next given that one necessarily has $\iota_K \Psi=-\tilde{K}\wedge \Psi$ and ${\cal L}_K\Psi=0$ it is possible to show that
\begin{align}
(-1)^{\text{deg}(\Psi)}d_H\left(\frac{e^{-\Phi}}{ 2K^2}\Psi K\right)&=\frac{1}{ 2K^2}(-1)^{\text{deg}(\Psi)}d_H\left(e^{-\Phi}\Psi\right) K+ e^{-\Phi}d\left(\frac{\tilde {K}+K}{2 K^2}\right)\wedge \Psi,\nn\\
d_H\left(\frac{e^{-\Phi}}{ 2K^2}K\Psi \right)&=-\frac{1}{ 2K^2}K d_H\left(e^{-\Phi}\Psi \right)-e^{-\Phi}d\left(\frac{\tilde {K}-K}{2 K^2}\right)\wedge \Psi,
\end{align}
allowing one to commute $K$ past $d_H$. Inside the pairings these terms simplify further due to 
\beq
K_1 K_2+K_2 K_1 \propto \mathbb{I},~~~~  d_H(e^{-\Phi}\Psi)=- (\iota_{K}+\tilde{K}\wedge )F=2(K_1 F-(-1)^{\deg{\Psi}}F K_2  ),~~~~F\epsilon_2=\overline{\epsilon}_1F=0.
\eeq
This ensures that inside the pairings the  $d_H(e^{-\Phi}\Psi)K$ terms vanish as
\beq
(K_1 F-(-1)^{\deg{\Psi}}F K_2  )K K_1 \epsilon_2\propto (K_1 F-(-1)^{\deg{\Psi}}F K_2  )\epsilon_2=0
\eeq
and the $K d_H(e^{-\Phi}\Psi)$ term as 
\beq
\overline{\epsilon}_1 K_2 K(K_1 F-(-1)^{\deg{\Psi}}F K_2)=0.
\eeq
The pairing conditions for the time-like case can thus be expressed as
\begin{subequations}
\begin{align}
\left( K\Psi K,~\Gamma^{MN}\bigg[d\left(\frac{\tilde K+K}{2K^2}\right)\wedge \Psi+\frac{1}{8}{\cal L}_{ \tilde{K}}\left(\frac{1}{K^2}\right)\Psi-e^{\Phi}F\bigg]\right)=0,\label{eq:TLpairing1}\\
\left( K\Psi K,~\bigg[d\left(\frac{\tilde K-K}{2K^2}\right)\wedge \Psi -\frac{1}{8}{\cal L}_{ \tilde{K}}\left(\frac{1}{K^2}\right)\Psi+e^{\Phi}F\bigg]\Gamma^{MN}\right)=0,\label{eq:TLpairing2}
\end{align}
\end{subequations}
in full generality.

\section{\texorpdfstring{Derivation of conditions for ${\cal N}=1$ AdS$_2$ in Type II supergravity}{Derivation of conditions for ${\cal N}=1$ AdS(2) in Type II supergravity}}\label{sec:derivation}
We take the most general Ansatz for a ten-dimensional solution with a warped AdS$_2$ factor, namely
\begin{align}
ds^2&=e^{2A}ds^2(\text{AdS}_2)+ds^2(\text{M}_8), \\F&= f_{\pm}+e^{2A}\text{vol}(\text{AdS}_2)\wedge\star_8 \lambda(f_\pm), ~~~~H=e^{2A}\text{vol}(\text{AdS}_2)\wedge H_1+H_3 
\end{align}
with the dilaton $\Phi$, warp factor $A$ and the forms $f_{\pm}$, $H_1$ and $H_3$ depending on M$_8$ directions only; here and elsewhere the upper signs are in Type IIA and the lower in Type IIB supergravity. The $d=10$ Majorana--Weyl Killing spinors are
\beq
\epsilon_{1}=\zeta_+\otimes\chi_{1+}+\zeta_-\otimes\chi_{1-},~~~~ \epsilon_{2}=\zeta_+\otimes\chi_{2\mp}+\zeta_-\otimes\chi_{2\pm},\label{eq:deq10spinor}
\eeq
where every spinor here is Majorana, $(\zeta_+,\zeta_-)$ are chiral Killing spinors on AdS$_2$ obeying \eqref{eq:ADS2KSE} and $(\chi_{i+},\chi_{i-})$ are chiral spinors on M$_8$. None of these latter spinors can vanish when $m\neq 0$; to see this one can consider the necessary spinorial supersymmetry conditions
\beq\label{eq:nozerointernalspinors}
(\nabla-\frac{1}{4}H-d\Phi)\epsilon_1=(\nabla+\frac{1}{4}H-d\Phi)\epsilon_2=0,
\eeq
given the AdS$_2$ Killing spinor equations maps between $\zeta_{+}$ and $\zeta_-$ under $\nabla$, we conclude that the only way to fix one of $(\chi_{i+},\chi_{i-})$ to zero is to also fix the inverse AdS$_2$ radius $m=0$, resulting in Mink$_2$.
We decompose the $d=10$ gamma matrices as
\beq
\Gamma_{\mu}= e^{A}\gamma^{(2)}_{\mu}\otimes \hat\gamma,~~~~\Gamma_a= \mathbb{I}\otimes \gamma_a,~~~~B^{(10)}=\mathbb{I}\otimes B.\label{eq:gammas8d}
\eeq
From this we can immediately compute $K$, $\tilde K$ 
\begin{align}
K&=\frac{1}{64}\bigg(e^{A}(|\chi_1|^2+|\chi_2|^2)v+e^{A}(\chi_1^{\dag}\hat\gamma \chi_1\mp\chi_2^{\dag}\hat\gamma \chi_2)u\bigg)-\frac{1}{32}f k,\label{eq:gendeq10Killingvector}\\
\tilde{K}&=\frac{1}{64}\bigg(e^{A}(|\chi_1|^2-|\chi_2|^2)v+e^{A}(\chi_1^{\dag}\hat\gamma \chi_1\pm\chi_2^{\dag}\hat\gamma \chi_2)u\bigg)-\frac{1}{32}f\tilde{k}
\end{align}
where $(f,u,v)$ are objects on AdS$_2$ defined in appendix \ref{sec:appenixAdS2bilinears}, we introduce $\chi_{1,2}:= \chi_{1,2+}+\chi_{1,2-}$, and
\beq
k_a:= \frac{1}{2}(\chi^{\dag}_{1}\gamma_a \chi_{1}\mp\chi^{\dag}_{2}\gamma_a \chi_{2}),~~~~\tilde{k}_a:= \frac{1}{2}(\chi^{\dag}_{1}\gamma_a \chi_{1}\pm\chi^{\dag}_{2}\gamma_a \chi_{2}).\label{eq:startargument}
\eeq
The next step is to use these bilinears to reduce \eqref{eq:10dsusy1} to conditions on M$_8$ only and, in the process, we will establish that \eqref{eq:noads3cond} is necessary for solutions for which AdS$_2$ does not only appear as a factor within a higher-dimensional  AdS space. 

First off $\nabla_{(M} K_{N)}=0$ imposes
\begin{align}
&\nabla_{(a}k_{b)}=0,\nn\\
&{\cal L}_{k}A+\frac{m}{2} e^{-A}(\chi_1^{\dag}\hat\gamma \chi_1\mp\chi_2^{\dag}\hat\gamma \chi_2)=0 ,\nn\\
&d(e^{-A}(|\chi_1|^2+|\chi_2|^2))=0,\nn\\
&d(e^{-A}(\chi_1^{\dag}\hat\gamma \chi_1\mp\chi_2^{\dag}\hat\gamma \chi_2))+2m e^{-2A}k=0.
\end{align}
These together imply that if $k$ is non-trivial for AdS$_2$ (i.e.\ $m\neq0$), then it's an isometry of the internal metric but not of the warp factor. Indeed one can introduce a local coordinate $\rho$ such that
\beq
-e^{-A}(\chi_1^{\dag}\hat\gamma \chi_1\mp\chi_2^{\dag}\hat\gamma \chi_2)=h(\rho),~~~~k=\frac{1}{2m}h'e^{2A}d\rho.
\eeq
Clearly $\rho$ spans this isometry, and $h$ parametrises diffeomorphism invariance in this direction. We can thus parametrise
\beq
k^a\partial_{x^{a}}=\partial_{\rho}, ~~ \Rightarrow ~~ |k|^2= \frac{1}{2m}h' e^{2A},
\eeq
allowing us to define the vielbein direction
\beq
\e^{k}:= \frac{k}{|k|}= \sqrt{\frac{h'}{2m}} e^{A} d\rho,~~~~ds^2(\text{M}_8)= (\e^{k})^2+ds^2(\text{M}_7).
\eeq
We demand that locally the warp factor is independent of $\rho$ since M$_8$ respects the isometry, which leads to
\beq
e^{A}= \sqrt{\frac{2m}{h'}} e^{A_7},~~~~ \partial_{\rho}A_7=0.
\eeq
Substituting this into the Lie derivative conditions then yields, without loss of generality,
\beq
h=\frac{2}{m} \tanh (\rho)
\eeq
and so the metric becomes
\beq
ds^2= e^{2A_7}\bigg[m^2\cosh^2\rho ds^2(\text{AdS}_2)+ d\rho^2\bigg]+ ds^2(\text{M}_7),\label{eq:endargument}
\eeq
which is AdS$_3$ rather than AdS$_2$.

Second, from $d\tilde{K}=\iota_{K}H$ we get
\begin{align}
&d\tilde{k}=\iota_{k}H_3,\label{eq:drildeKa}\\
&\iota_{k}(e^{2A}H_1)= m e^{A}(|\chi_1|^2-|\chi_2|^2),\label{eq:drildeKb}\\
&e^{-A}(\chi_1^{\dag}\hat\gamma \chi^1\mp \chi^{\dag}_{2}\hat\gamma\chi_2)(e^{2A}H_1)+d(e^{A}(|\chi_1|^2-|\chi_2|^2))=0,\label{eq:drildeKc}\\
&e^{-A}(|\chi_1|^2+|\chi_2|^2)(e^{2A}H_1)+d(e^{A}(\chi_1^{\dag}\hat\gamma \chi^1\pm \chi^{\dag}_{2}\hat\gamma\chi_2))=2m\tilde{k}\label{eq:drildeKd}.
\end{align}
Proceeding with the local coordinate $\rho$ again, the fact that away from source $d(e^{2A}H_1)=0$ and given  \eqref{eq:drildeKb}-\eqref{eq:drildeKc} we have
\beq
e^{A}(|\chi_1|^2-|\chi_2|^2)=g(\rho),~~~~m\partial_{\rho}\log g=h ~~ \Rightarrow ~~ e^{2A}H_1= b m \cosh^2\rho d\rho,~~~~db=0,
\eeq
then \eqref{eq:drildeKd} implies
\beq
d\tilde k=0, ~~ \Rightarrow ~~ \iota_{k}H_3=0, ~~ \Rightarrow ~~ {\cal L}_{k}H_3= \iota_{k}dH_3
\eeq
so when $dH_3=0$, i.e.\ away from sources, the NSNS flux also respects the isometries of AdS$_3$ if $k \neq 0$ (at least locally away from sources for $d(e^{2A}H_1)$). 

Third, one can show that $\Psi$ decomposes as
\begin{align}
\Psi= &\pm\frac{1}{2} f\bigg(\psi_{\pm}-e^{2A}\text{vol}(\text{AdS}_2)\wedge\hat\psi_{\pm} \bigg)\mp\frac{e^{A}}{2}\bigg(v\wedge\psi_{\mp}+u\wedge\hat \psi_{\mp}\bigg), 
\end{align}
where we define the $d=8$ bispinors
\beq
\psi:= \chi_1\otimes \chi_2^{\dag},~~~~\hat\psi:= \hat\gamma\chi_1\otimes \chi_2^{\dag},
\eeq
and the $\pm$ subscripts refer to the even/odd form degree parts of these.
On the other hand we find 
\begin{align}
(\iota_{K}+\tilde K\wedge)f_{\pm}&=\frac{1}{16}\bigg[(\zeta_+\otimes \overline{\zeta}_-)_2\wedge(\iota_{k}+\tilde{k}\wedge)\star_8(e^{2A}\lambda f_{\pm})-(\zeta_+\otimes \overline{\zeta}_-)_0(\iota_{k}+\tilde{k}\wedge)f_{\pm}\nn\\
&+\frac{1}{4}v\wedge \bigg(e^{-A}(\chi_1^{\dag}\hat\gamma \chi^1\mp \chi^{\dag}_{2}\hat\gamma\chi_2)\star_8(e^{2A} \lambda f_{\pm})+e^{A}(|\chi_1|^2-|\chi_2|^2)f_{\pm}\bigg)\nn\\
&+\frac{1}{4}u\wedge \bigg(e^{-A}(|\chi_1|^2+|\chi_2|^2)\star_8\lambda (e^{2A} f_{\pm})+e^{A}(\chi_1^{\dag}\hat\gamma \chi_1\pm \chi^{\dag}_{2}\hat\gamma\chi_2)f_{\pm}\bigg)\bigg].
\end{align}
So we find 
\begin{align}
&d_{H_3}(e^{-\Phi}\psi_{\pm})=\pm \frac{1}{16}(\iota_{k}+\tilde{k}\wedge)f_{\pm},\label{eq:bicondsa}\\
&d_{H_3}(e^{A-\Phi}\psi_{\mp})=\mp\frac{1}{32}\bigg(e^{-A}(\chi_1^{\dag}\hat\gamma \chi^1\mp \chi^{\dag}_{2}\hat\gamma\chi_2)\star_8\lambda (e^{2A} f_{\pm})+e^{A-\Phi}(|\chi_1|^2-|\chi_2|^2)f_{\pm}\bigg),\label{eq:bicondsb}\\
&d_{H_3}(e^{A-\Phi}\hat\psi_{\mp})- m e^{-\Phi}\psi_{\pm}=\mp\frac{1}{32}\bigg(e^{-A}(|\chi_1|^2+|\chi_2|^2)\star_8\lambda(e^{2A} f_{\pm})+e^{A}(\chi_1^{\dag}\hat\gamma \chi_1\pm \chi^{\dag}_{2}\hat\gamma\chi_2)f_{\pm}\bigg),\label{eq:bicondsc}\\
&d_{H_3}(e^{2A-\Phi}\hat\psi_{\pm})+e^{2A-\Phi}H_1\wedge\psi_{\pm}-2m e^{A-\Phi}\psi_{\mp}=\mp \frac{1}{16}(\iota_{k}+\tilde{k}\wedge)\star_8 \lambda (e^{2A} f_{\pm}),\label{eq:bicondsd}
\end{align}
Returning to the local coordinate $\rho$, and given that away from sources the RR flux should obey
\beq
d_{H_3}f_{\pm}=d_{H_3}(e^{2A}\star_8\lambda f_{\pm})-e^{2A}H_1\wedge f_{\pm}=0,
\eeq
taking $d_{H_3}$\eqref{eq:bicondsb} yields
\beq
d_{H_3}(-h (e^{2A}\star_8 \lambda f_{\pm})+g f_{\pm})=  -h'd\rho \wedge (e^{2A}\star_8 \lambda f_{\pm})=0,
\eeq
i.e. $f_{\pm}$ is orthogonal to $k$, so what has been derived thus far implies
\beq
{\cal L}_{k} f_{\pm}=0,
\eeq
so the RR fluxes also respect the isometries of AdS$_3$ away from possible sources along $\rho$. We  conclude that in any regular region of a solution
\beq
k \neq 0 ~~ \Rightarrow ~~ \text{warped AdS}_3.
\eeq
Note that a similar result was found for AdS$_3$ solutions preserving ${\cal N}=(1,1)$ supersymmetry in \cite{Macpherson:2021lbr}. This argument of course breaks down if $m=0$, the Mink$_2$ limit: For this we generically have ${\cal L}_{k}A=0$ and one can show that, when non-zero,  $k^a\partial_{x^{a}}$ is actually a Killing vector with respect to the entire solution under which the spinors are uncharged. There is obviously the potential for Mink$_2$ to get enhanced to Mink$_3$ in this case, but this is no longer guaranteed as $k$ can appear fibered over the rest of the internal space. One is still free to fix $k=0$ when $m=0$, the point is that it is no longer general to do so --- for $m\neq 0$ there is no such problem. Thus if we are interested in AdS$_2$ solutions, we should fix
\begin{align}
k&=0,~~~~ \chi_1^{\dag}\hat\gamma \chi_1=\pm\chi_2^{\dag}\hat\gamma \chi_2,~~~~ |\chi_1|^2=|\chi_2|^2= c e^{ A},\nn\\
d\tilde k&=0,~~~~2c e^{2A}H_1=2m\tilde{k}-d(e^{A}(\chi_1^{\dag}\hat\gamma \chi_{1}\pm \chi^{\dag}_{2}\hat\gamma\chi_2)),\label{eq: AdS2cond}
\end{align}
for $c>0$ some constant, to solve all these conditions. 

This reduces the $d=10$ 1-forms $K$ to
\begin{align}
K&=\frac{1}{32}e^{2A}c v, ~~ \Rightarrow ~~ (K)^{\mu}=\frac{c}{32}v^{\mu},~~~~ K^2=-\left(\frac{c}{32}\right)^2 e^{2A}f^2,
\end{align}
so the ten-dimensional Killing vector is  time-like for all AdS$_2$ solutions that are not merely the embedding of AdS$_2$ into AdS$_3$. The bispinor conditions then truncate to 
\begin{align}
&d_{H_3}(e^{-\Phi}\psi_{\pm})=\pm \frac{1}{16}\tilde{k}\wedge f_{\pm},\label{eq:bicondsAdS2a}\\
&d_{H_3}(e^{A-\Phi}\psi_{\mp})=0,\label{eq:bicondsAdS2b}\\
&d_{H_3}(e^{A-\Phi}\hat\psi_{\mp})- m e^{-\Phi}\psi_{\pm}=\mp\frac{c }{16}e^{2A}\star_8\lambda f_{\pm}\mp\frac{1}{32}e^{A}(\chi_1^{\dag}\hat\gamma \chi_1\pm \chi^{\dag}_{2}\hat\gamma\chi_2)f_{\pm},\label{eq:bicondsAdS2c}\\
&d_{H_3}(e^{2A-\Phi}\hat\psi_{\pm})+e^{2A-\Phi}H_1\wedge\psi_{\pm}-2m e^{A-\Phi}\psi_{\mp}=\mp \frac{1}{16}\tilde{k}\wedge\star_8 e^{2A}\lambda  f_{\pm}.\label{eq:bicondsAdS2d}
\end{align}
Note that the Bianchi identities of $H_3$ and $f_{\pm}$ imply that of $e^{2A}\star_8\lambda(f_{\pm})$ by \eqref{eq:bicondsAdS2a} and $d_{H_3}$\eqref{eq:bicondsAdS2c}. Further a supersymmetric solution must obey $(\tilde{K}\wedge +\iota_{K})\Psi=(K\wedge -\iota_{\tilde{K}})\Psi=0$, which in terms of eight-dimensional conditions becomes
\begin{align}
e^{A}c \psi_{\mp}&=\tilde{k}\wedge \psi_{\pm},~~~~e^{A}c {\cal G} \psi_{\mp}=\tilde{k}\wedge \hat\psi_{\pm},~~~~e^{A}c {\cal G}\psi_{\pm}=\tilde{k}\wedge\hat\psi_{\mp}+e^{A}c \hat\psi_{\pm},\nn\\
e^{A}c\hat\psi_{\mp}&=-\iota_{\tilde{k}}\hat\psi_{\pm},~~~~e^{A}c{\cal G}\hat\psi_{\mp}=-\iota_{\tilde{k}}\psi_{\pm},~~~~e^{A}c{\cal G}\hat\psi_{\pm}=-\iota_{\tilde{k}}\psi_{\mp}+e^{A}c\psi_{\pm}
\end{align}
where we employ the short-hand $e^{A}c{\cal G} =\chi_1^{\dag}\hat\gamma \chi_1=\pm\chi_2^{\dag}\hat\gamma \chi_2$. 
The first line of which, along with \eqref{eq: AdS2cond}, can be used to prove the redundancy of \eqref{eq:bicondsAdS2b} and \eqref{eq:bicondsAdS2d}.

What remains is to reduce the pairing constraints to $d=8$ conditions. Going forward, and in the main text we define
\beq
{\cal G}= \cos\beta,~~~~\tilde{k}= c e^{A} \sin\beta V,
\eeq
for $V$ a unit norm 1-form, as we are free to do without loss of generality. Note that it is not possible to fix $\sin\beta=0$ as that would set some of the chiral $d=8$ spinors to zero and result in $\text{Mink}_2$ as explained around \eqref{eq:nozerointernalspinors}. 
Taking \eqref{eq:TLpairing1}-\eqref{eq:TLpairing2} as our starting point, for the case at hand we have
\beq
K=\frac{c}{32} e^{2A} f^2 dt,~~~~ K^2=  -\left(\frac{c}{32}\right)^2 e^{2A}f^2,~~~~\tilde{K}= \frac{c f e^A}{32}\bigg(- \cos\beta \e^r+ \sin\beta V\bigg),
\eeq
where $\e^t= e^{mr}dt$, $\e^r= dr$, from which it follows that
\beq
d\left(\frac{\tilde K\pm K}{2K^2}\right)\wedge\Psi= \frac{1}{2}\bigg[\frac{1}{K^2}(\iota_K H)\wedge \Psi-  d\left(\frac{1}{K^2}\right)\wedge \iota_K\Psi\bigg]= \frac{16}{e^A f c}( \e^r\wedge H_1\wedge \Psi-2(dA+m e^{-A}\e^r)\wedge\iota_{\e^t} \Psi)
\eeq
where we have used that $\tilde{K}\wedge \Psi=- \iota_K\Psi$. We also have that
\beq
K\Psi K \propto \Gamma^0\Psi \Gamma^0,~~~~\frac{1}{8}{\cal L}_{ \tilde{K}}\left(\frac{1}{K^2}\right)=-\frac{8}{c e^A f}( \cos\beta  m e^{-A}+\sin\beta {\cal L}_{V} dA).
\eeq
We are now ready to simplify \eqref{eq:bicondsAdS2b}-\eqref{eq:bicondsAdS2d} component by component, we find it useful to  split the $d=10$ index as $M=(t,r,a)$, i.e.\ the two  directions along AdS$_2$ and along M$_8$. In what follows we make frequent use of \eqref{eq:pairingidenities}.

The $tr$ components are both equal and give rise to
\beq\label{eq:8formpairing}
(\psi_{\pm},f_{\pm})_8=\pm\frac{c}{4} e^{-\Phi}\left(m-\frac{1}{2}e^A\sin\beta \iota_{V} H_1\right)\text{vol}(\text{M}_8).
\eeq
The $ta$ components yield
\beq
(\psi_{\mp},\star\lambda f_{\pm})_7=0,~~~~(\hat\psi_{\mp}, f_{\pm})_7=\pm \frac{1}{8}e^{A-\Phi}c\star_8( 2dA+\cos\beta H_1).
\eeq
The $ra$ components yield
\beq
(\psi_{\mp}, f_{\pm})_7=0,~~~~(\hat\psi_{\mp}, \star\lambda f_{\pm })_7=\pm \frac{1}{8}e^{A-\Phi}c\star_8( 2 \cos\beta dA+ H_1-2 e^{-A} m \sin\beta V),
\eeq
by wedging these conditions with $V$ one can form a linear combination that implies $\sin\beta$\eqref{eq:8formpairing} $=0$, but one cannot fix $\sin\beta=0$ hence the 8-form pairing is implied by the $d=7$ ones.
The $ab$ components are more complicated, but through a lengthy computation, extracting all the conditions that the 7-form pairings impose on the flux, it is possible to show that they too are implied. By manipulating ($\hat\Psi_{\pm}$,\eqref{eq:bicondsAdS2a})$_7$  or  ($\Psi_{\mp}$,\eqref{eq:bicondsAdS2c})$_7$ one extracts
\beq
(\psi_{\mp},\star\lambda f_{\pm}+ \cos\beta f_{\pm})_7=0,
\eeq
from which it follows that $(\psi_{\mp},\star\lambda f_{\pm})_7=0$ is implied by $(\psi_{\mp}, f_{\pm})_7=0$.

We have now extracted necessary and sufficient $d=8$ geometric conditions for minimally supersymmetric AdS$_2$ solutions of Type II supergravity, our results are summarised in section \ref{sec:SUSYads2}.

\section{\texorpdfstring{Comment on \cite{Hong:2019wyi} and ${\cal N}=1$ AdS$_2$ in $d=11$ supergravity}{Comment on \cite{Hong:2019wyi} and ${\cal N}=1$ AdS(2) in $d=11$ supergravity}}\label{sec:comment}
In \cite{Hong:2019wyi} a class of supersymmetic AdS$_2$ solutions in $d=11$ supergravity was derived under the assumption that the internal 9-manifold supports an SU$(4)$-structure. In this appendix we show that it is actually general, as solutions supporting any other structure are merely the  embeddings of AdS$_2$ into supersymmetic AdS$_3$ solutions.

In \cite{Gauntlett:2002fz}, geometric conditions that are necessary  for supersymmetry in $d=11$ supergravity were derived; they are defined in terms of the following bilinears
\beq
K^{(11)}:=\epsilon^{\dag}\Gamma_0\Gamma_{\underline M} \epsilon \, \e^{\, \underline{M}},~~~~ \Omega^{(11)}:=\frac{1}{2}\epsilon^{\dag}\Gamma_0\Gamma_{\underline{M}
	\underline{N}}\epsilon \, \e^{\, \underline{M}\underline{N}} ,~~~~\Sigma^{(11)}:=\frac{1}{5!}\epsilon^{\dag}\Gamma_0\Gamma_{\underline{M}_1\dots\underline{M}_5}\epsilon \, \e^{\, \underline{M}_1\dots\underline{M}_5},\label{eq:deq11forms}
\eeq
where $\epsilon$ is the Majorana Killing spinor of $d=11$ supergravity, $\Gamma_{\underline{M}}$ are flat spacetime gamma matrices in $d=11$ and $\e^{\, \underline{M}}$ a corresponding vielbein. The conditions these objects must satisfy are
\begin{align}
d\Omega^{(11)}&=\iota_{K^{(11)}} G,~~~~\nabla_{(M} K^{(11)}_{N)}=0,\nn\\
d\Sigma^{(11)}&= \iota_{K^{(11)}} \star_{11}G- \Omega^{(11)}\wedge G,\nn\\
\star_{11} dK^{(11)}&=\frac{2}{3}\Omega^{(11)}\wedge \star_{11}G-\frac{1}{3}\Sigma^{(11)}\wedge G.\label{eq:deq11susy}
\end{align}
Clearly this makes $(K^{(11)})^{M}\partial_M$ a Killing vector, and it turns out that when this is taken to be timelike the above system is also sufficient for supersymmetry.

We would now like to use \eqref{eq:deq11susy} to establish that, for AdS$_2$ solutions, it is necessary for the internal space to support an SU(4)-structure. To this end we decompose the metric and fluxes as
\beq
ds^2=  e^{2\Delta} ds^2(\text{AdS}_2)+ ds^2(\hat{\text{M}}_9),~~~~G=  e^{2\Delta}\text{vol}(\text{AdS}_2)\wedge G_2+ G_4, 
\eeq
and take our $d=11$ gamma matrices to decompose as
\beq
\Gamma_{\mu}= e^{\Delta}\gamma^{(2)}_{\mu}\otimes \mathbb{I},~~~~\Gamma_{a}=\sigma_3\otimes \gamma^{(9)}_a,~~~~a=1,\dots,9.
\eeq
We will define the $d=9$ gamma matrices in terms of the $d=8$ ones in \eqref{eq:gammas8d} as $\gamma^{(9)}_a=\gamma_a$ for $a=1,\dots,8$  and $\gamma^{(9)}_9=\gamma_{1\dots 8}$, in this way the intertwiner defining $d=11$ Majorana conjugation is the same as it was in $d=10$ and we can take the $d=11$ spinor to be
\beq
\epsilon= \epsilon_1+ \epsilon_2,
\eeq
where $\epsilon_{1,2}$ are defined as in \eqref{eq:deq10spinor} (for the upper signs specifically). We now compute the 1-form in \eqref{eq:deq11forms} and find
\begin{align}
K^{(11)}&=-\left( e^{\Delta}(|\chi_1|^2+|\chi_2|^2)v+ e^{\Delta}(\chi_1^{\dag}\gamma^{(9)}_9\chi_1-\chi_2^{\dag}\gamma^{(9)}_9\chi_2)u -f \xi\right),\nn\\
\xi&= -(\chi_{1+}^{\dag}\gamma^{(9)}_a\chi_{2-}+\chi_{1+}^{\dag}\gamma^{(9)}_a\chi_{2-}+2(\chi_{1+}^{\dag}\gamma^{(9)}_a\chi_{2+}+\chi_{2-}^{\dag}\gamma^{(9)}_a\chi_{1-}))\e^a
\end{align}
where $(f,u,v)$ are the scalar and two 1-forms on AdS$_2$ defined in appendix \ref{sec:appenixAdS2bilinears}. We note that this has precisely the same structure as  \eqref{eq:gendeq10Killingvector}, so  imposing $\nabla_{(M} K^{(11)}_{N)}=0$ leads to a repeat of the argument between  \eqref{eq:startargument} and  \eqref{eq:endargument}; jumping to the punch line:  generic AdS$_2$ solutions experience an enhancement to  AdS$_3$, for true AdS$_2$ solutions we must impose
\beq
d(e^{- \Delta}(|\chi_1|^2+|\chi_2|^2))=0,~~~~\chi_1^{\dag}\gamma^{(9)}_9\chi_1=\chi_2^{\dag}\gamma^{(9)}_9\chi_2,~~~~   \xi = 0,
\eeq
which in turn makes $(K^{(11)})^{M}\partial_M$ necessarily timelike, just as the analogue was for Type II supergravity.
Solving these conditions amounts to imposing that
\beq
\chi_{1+}=\chi_{2+}=0,~~~~|\chi_1|^2=|\chi_2|^2=\frac{1}{2}e^{\Delta}
\eeq
without loss of generality. Then upon defining a real vector $V^{(4)}$ and SU(4)-structure forms $(J^{(4)},\Omega^{(4)})$ as
\beq
e^{\Delta} V^{(4)}= \chi^{\dag}\gamma^{(9)}_{\underline{a}} \chi \, \e^{\,\underline{a}},~~~~ e^{\Delta}J^{(4)}=-\frac{i}{2}\chi^{\dag}\gamma^{(9)}_{{\underline{a}}{\underline{b}}} \chi \, \e^{\, {\underline{a}}{\underline{b}}},~~~~e^{\Delta}\Omega^{(4)}=\frac{1}{4!}\chi^{c\dag}\gamma^{(9)}_{{\underline{a}}{\underline{b}}{\underline{c}}{\underline{d}}} \chi \, \e^{{\,\underline{a}}{\underline{b}}{\underline{c}}{\underline{d}}},
\eeq
for
\beq
\chi:= i(\chi_1+ i \chi_2).
\eeq
We find that the forms in \eqref{eq:deq11forms} become
\begin{align}
K^{(11)}&=-e^{2\Delta}v,~~~~ \Omega^{(11)}=e^{\Delta}(- u\wedge V^{(4)}- f J^{(4)}),\\
\Sigma^{(11)}&= e^{\Delta}\left(-e^{2\Delta} f \text{vol}(\text{AdS}_2)\wedge J^{(4)}\wedge V^{(4)}+\frac{1}{2}e^{\Delta} v\wedge J^{(4)}\wedge J^{(4)}-e^{\Delta} u \wedge \text{Re}\Omega^{(4)}+ f V^{(4)}\wedge \text{Im}\Omega^{(4)}\right)\nn,
\end{align}
which are clearly spanned by forms on AdS$_2$ and those defining an SU(4)-structure in $d=9$. These precisely reproduce the equivalent forms of \cite{Hong:2019wyi}, so we have proved that the SU(4)-structure ``assumption'' of that paper, is actually no assumption at all but rather a necessary condition for true supersymmetic AdS$_2$ solutions in $d=11$ supergravity. Fixing $m=1$ as \cite{Hong:2019wyi} does, for completeness, we quote the necessary and sufficient geometric conditions for supersymmetry in $d=11$
\begin{subequations}
\begin{align}
&d(e^{\Delta} J^{(4)})=0,\label{eq:deq11cond1}\\
&d(e^{2\Delta}V^{(4)}) + e^{\Delta}J^{(4)}+ e^{2\Delta} G_2=0,\label{eq:deq11cond2}\\
&d(e^{\Delta} V^{(4)}\wedge \text{Im}\Omega^{(4)})-e^{\Delta} J^{(4)}\wedge G_4=0,\label{eq:deq11cond3}\\
&d(e^{2\Delta} \text{Re}\Omega^{(4)})- e^{\Delta} V\wedge \text{Im}\Omega^{(4)}+ e^{2\Delta}(\star_9 G_4-V^{(4)}\wedge G_4)=0,\label{eq:deq11cond6}\\
&\star_9\big(2V^{(4)}\wedge \star_9 G_2+  \text{Re}\Omega^{(4)}\wedge G_4\big)+6 d\Delta=0,\label{eq:deq11cond4}\\
&J^{(4)}\wedge J^{(4)}\wedge G_4=0,\label{eq:deq11cond5}\\
&e^{\Delta}\big(2J^{(4)}\wedge \star_9 G_2-  V\wedge \text{Im}\Omega^{(4)}\wedge G_4\big)=6 \text{Vol}(\text{M}_9)\label{eq:deq11cond8}.
\end{align}
\end{subequations}
In \cite{Hong:2019wyi} the condition
\beq
V^{(4)}\wedge (\text{Im}\Omega^{(4)} \wedge G_2+ J^{(4)}\wedge G_4)=0,\label{eq:algebcond7}
\eeq
is also quoted, but this is implied by \eqref{eq:deq11cond2} and \eqref{eq:deq11cond3}.

\end{document}